\documentclass[preprint,12pt,review]{elsarticle}

\usepackage{amssymb}
\usepackage{amsmath}
\usepackage{float}
\usepackage{xcolor}

\usepackage{breqn}
\usepackage{siunitx}
\usepackage{tikz}
\usepackage{pgfkeys}
\usepackage{setspace}
\usetikzlibrary{calc,patterns,angles,quotes}

\usepackage{lineno}

\usepackage{subcaption}
\usepackage{pdfpages}

\usepackage{hyperref}
\usepackage{comment}

\graphicspath{{figures/}{figures/dEdx/}}

\journal{Nuclear Instruments and Methods}

\begin{document}

\begin{frontmatter}





\title{Analysis of test beam data taken with a prototype of TPC with resistive Micromegas  for the T2K Near Detector upgrade}

\author[saclay]{D.~Atti\'e}
\author[ifae]{O.~Ballester}

\author[ifj]{M.~Batkiewicz-Kwasniak}
\author[lpnhe]{P.~Billoir}
\author[lpnhe]{A.~Blanchet}
\author[lpnhe]{A.~Blondel}
\author[saclay]{S.~Bolognesi}
\author[saclay]{R.~Boullon}
\author[saclay]{D.~Calvet}
\author[ifae,pilar]{M.~P.~Casado}

\author[bari]{M.G.~Catanesi}
\author[legnaro]{M.~Cicerchia}
\author[padova]{G.~Cogo}
\author[saclay]{P.~Colas}
\author[padova]{G.~Collazuol}
\author[lpnhe]{C.~Dalmazzone}
\author[saclay]{T.~Daret}
\author[saclay]{A.~Delbart}
\author[ifae,annalisa]{A.~De Lorenzis}
\author[cern]{S.~Dolan}

\author[wut]{K.~Dygnarowicz}
\author[lpnhe]{J.~Dumarchez}
\author[saclay]{S.~Emery-Schrenk}
\author[saclay]{A.~Ershova}
\author[saclay]{G.~Eurin}
\author[padova]{M.~Feltre}
\author[padova]{C.~Forza}
\author[padova]{L.~Giannessi}
\author[lpnhe]{C.~Giganti}
\author[legnaro]{F.~Gramegna}
\author[padova]{M.~Grassi}
\author[lpnhe]{M.~Guigue}
\author[aachen]{P.~Hamacher-Baumann}
\author[saclay]{S.~Hassani}
\author[saclay]{D.~Henaff}

\author[padova]{F.~Iacob}
\author[ifae]{C.~Jes\'{u}s-Valls}
\author[saclay]{S.~Joshi}

\author[wut]{R.~Kurjata}
\author[padova]{M.~Lamoureux}
\author[napoli]{A.~Langella}
\author[saclay]{J.~F.~Laporte}

\author[napoli]{L.~Lavitola}
\author[saclay]{M.~Lehuraux}
\author[padova]{A.~Longhin}
\author[ifae]{T.~Lux}
\author[bari]{L.~Magaletti}
\author[legnaro]{T.~Marchi}
\author[lpnhe]{L.~Mellet}
\author[padova]{M.~Mezzetto}
\author[cern]{L.~Munteanu}
\author[lpnhe]{Q.~V.~Nguyen}
\author[lpnhe]{Y.~Orain}
\author[padova]{M.~Pari}
\author[lpnhe]{J.-M.~Parraud}
\author[bari]{C.~Pastore}
\author[padova]{A.~Pepato}
\author[lpnhe]{E.~Pierre}
\author[ifae]{C.~Pio Garcia}

\author[lpnhe]{B.~Popov}
\author[saclay]{J.~Porthault}
\author[ifj]{H.~Przybiliski}
\author[padova]{F.~Pupilli}

\author[aachen]{T.~Radermacher}
\author[bari]{E.~Radicioni}
\author[saclay]{F.~Rossi}
\author[aachen]{S.~Roth}
\author[lpnhe]{S.~Russo}

\author[wut]{A.~Rychter}
\author[padova]{L.~Scomparin}
\author[aachen]{D.~Smyczek}
\author[aachen]{J.~Steinmann}
\author[lpnhe]{S.~Suvorov}
\author[ifj]{J.~Swierblewski}
\author[lpnhe]{D.~Terront}
\author[aachen]{N.~Thamm}
\author[lpnhe]{F.~Toussenel}
\author[bari]{V. Valentino}
\author[ifae]{M.~Varghese}

\author[saclay]{G.~Vasseur}
\author[lpnhe]{U.~Virginet}

\author[lpnhe]{U.~Yevarouskaya\fnref{fnref1}}
\author[wut]{M.~Ziembicki}
\author[lpnhe]{M.~Zito}
%
\address[saclay]{IRFU, CEA, Universit\'e Paris-Saclay, Gif-sur-Yvette, France}
\address[ifj]{H. Niewodniczanski Institute of Nuclear Physics PAN, Cracow, Poland}
\address[lpnhe]{LPNHE, Sorbonne Universit\'e, CNRS/IN2P3, Paris, France}
\address[bari]{INFN sezione di Bari, Universit\`a di Bari e Politecnico di Bari, Italy}
\address[legnaro]{INFN: Laboratori Nazionali di Legnaro (LNL), Padova, Italy}
\address[padova]{INFN Sezione di Padova and Universit\`a di Padova, Dipartimento di Fisica e Astronomia, Padova, Italy}
\address[aachen]{RWTH Aachen University, III.~Physikalisches Institut, Aachen, Germany}
\address[ifae]{Institut de F\'isica d’Altes Energies (IFAE) - The Barcelona Institute of Science and Technology (BIST), Campus UAB, 08193 Bellaterra (Barcelona), Spain}
\address[wut]{Warsaw University of technology, Warsaw, Poland}
\address[cern]{CERN, European Organization for Nuclear Research, Geneva, Switzerland}

\address[napoli]{INFN Sezione di Napoli and Universit\`a di Napoli Federico II, Dipartimento di Fisica, Napoli, Italy}
\address[annalisa]{Qilimanjaro Quantum Tech, Barcelona 08007, Spain}
\address[pilar]{Departament de Física, Universitat Autònoma de Barcelona
}
\address[cern]{CERN, European Organization for Nuclear Research, Geneva, Switzerland}

\cortext[cor1]{Corresponding author}
\fntext[fnref1]{uyevarou@lpnhe.in2p3.fr}

\begin{abstract}

In this paper we describe the performance of a prototype of the High Angle Time Projection Chambers (HA-TPCs) that are being produced for the Near Detector (ND280) upgrade of the T2K experiment.
The two HA-TPCs of ND280 will be instrumented with eight Encapsulated Resistive Anode Micromegas (ERAM) on each endplate, for a total of 32 ERAMs. This innovative technique allows the detection of the charge emitted by ionization electrons over several pads, improving the determination of the track position.

The TPC prototype has been equipped with the first ERAM module produced for T2K and with the HA-TPC readout electronics chain and it has been exposed to an electron beam at DESY in order to measure spatial and dE/dx resolution. In this paper we characterize the performances of the ERAM and, for the first time, we compare them with a newly developed simulation of the detector response.

Spatial resolution better than 800 ${\rm\text{\textmu} m}$ and dE/dx resolution better than 10\% are observed for all the incident angles and for all the drift distances of interest. All the main features of the data are correctly reproduced by the simulation and these performances fully fulfill the requirements for the HA-TPCs of T2K.

\end{abstract}

\begin{keyword}
Resistive Micromegas, T2K Near Detector Time Projection Chambers



\end{keyword}

\end{frontmatter}

\tableofcontents
\newpage
\section{Introduction and physics motivations}

T2K (``Tokai-to-Kamioka")~\cite{Abe:2011ks} is a long-baseline neutrino oscillation experiment situated in Japan that has been taking data since 2010. By using an intense muon neutrino beam produced at the J-PARC accelerator complex and searching for the appearance of electron neutrinos at the far detector, Super-Kamiokande, T2K provided the first observation of muon to electron neutrino oscillations~\cite{Abe:2011sj, Abe:2013hdq}. Recently first hints of Charge-Parity (CP) violation in the leptonic sector were also published by T2K~\cite{Abe:2019vii}.

In order to confirm these hints, T2K is now preparing the second phase of the experiment, that includes an upgrade of the neutrino beamline \cite{T2K:2019eao} and of the off-axis Near Detector complex, ND280 \cite{T2K:2019bbb}. The ND280 is a multi-purpose detector with several sub-detectors installed inside the UA1/NOMAD magnet that provides a magnetic field of 0.2~T. The core of ND280 is a tracker system, composed by two Fine Grained Detectors (FGDs)~\cite{T2KND280FGD:2012umz} and three Time Projection Chambers (TPCs)~\cite{T2KND280TPC:2010nnd} instrumented with Bulk Micromegas modules~\cite{Giomataris:2004aa}. The TPCs are used to track charged particles emitted in neutrino interactions and to measure their charges and momenta as well as to perform particle identification based on the ionization energy losses in the gas. 
The ND280 has been extensively used in all T2K oscillation analyses and it allows for a reduction of systematic uncertainties to the level of 4--5\%~\cite{T2K:2021xwb}. These uncertainties mostly come from our limited knowledge of neutrino interactions with nuclei, and of the neutrino beam properties (energy spectrum and composition). 

An upgrade of ND280 is being constructed~\cite{Abe:2019whr}, with the goal of further reducing these systematic uncertainties~\cite{Dolan:2021hbw}. It consists in replacing one of the ND280 sub-detectors, the $\pi^0$ detector (P0D)~\cite{Assylbekov:2011sh}, with a new tracker system composed by a 3-dimensional scintillator target (Super-FGD)~\cite{Blondel:2020hml}, made of $\sim$2 millions scintillator cubes of 1~cm$^3$, each readout by three wavelength shifting fibers, two High Angle TPCs (HA-TPCs) and six Time-Of-Flight (TOF) planes~\cite{Korzenev:2021mny}. This upgrade will be installed at J-PARC in 2023. Among other improvements, the presence of the HA-TPCs will increase the reconstruction efficiency for the tracks produced in neutrino (antineutrino) interactions with nuclei, emitted at large angle or in backward direction with respect to the incoming neutrino.
The main goals of the two new HA-TPCs are reconstruction of charged particles trajectories, measurements of their momenta and particle identification.  These goals require a good spatial resolution and a precise ionization energy loss measurement.


Each endplate of the HA-TPC will be instrumented with 8 Encapsulated Resistive Anode Micromegas (ERAM)~\cite{Attie:2687703}.
The first ERAM detector produced for ND280 upgrade was initially tested using an X-ray test bench at CERN and then mounted on a prototype of the field cages that are being constructed for the HA-TPCs. This field cage prototype has the same construction materials, 
the same drift length and the same strip foils configuration to produce a uniform electric field as the cages that are being constructed for the ND280 upgrade. Also, the front-end electronics chain that will be used for the HA-TPCs, including two Front-End-Cards (FECs) each hosting eight Asic For TPC Electronic Readout (AFTER) chips~\cite{Baron:2008zza}, and one Front-End-Mezzanine (FEM), all equipped with their cooling plates, was mounted on the field cage. The FEM was connected to a Trigger and Data Concentrator Module (TDCM) \cite{Calvet:2018lac} used for the data transfer to a DAQ computer via a Modular Interactive Data Acquisition System (MIDAS)~\cite{MIDAS} front-end.

\begin{figure}[H]
    \centering
        \includegraphics[width=0.6\linewidth]{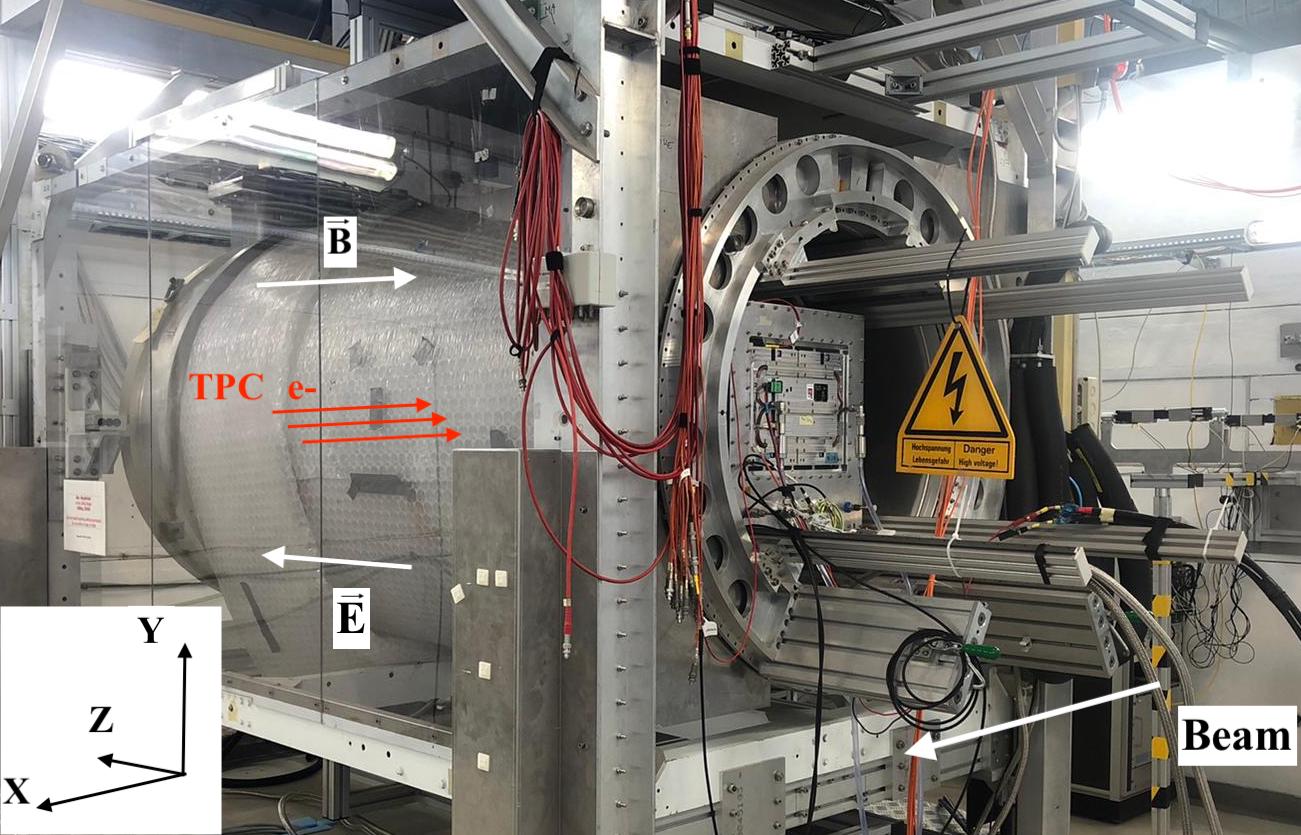} \
    \caption{The overview of the experimental setup: the HA-TPC prototype is placed inside the PCMAG of the DESY T24/1 facility.
}
    \label{fig:orient}
\end{figure}

The TPC prototype was placed at the DESY T24/1 facility~\cite{Diener:2018qap} inside a large-bore superconducting solenoid, called PCMAG, that provides a magnetic field of intensity up to 1.25~T and it was exposed to a beam of electrons with momenta between 1 and 4 GeV/\emph{c}, see Fig.\ref{fig:orient}. As we will show in this paper, this test beam campaign allowed us to validate the performances of the TPCs for tracks with different incident angles with respect to the ERAM detector and for all the drift distances of interest for the T2K TPCs. 

This paper uses similar methods as those described in Refs.~\cite{Attie:2019hua,Attie:2021yeh} 
to analyze the test beam data.
The main novelty of this paper is that the data are also compared with a simulation that has been developed by using the ND280 software~\cite{Abe:2011ks}, adding the HA-TPCs ERAM geometry and the features of the resistive layers and of the AFTER chip electronics response.



As it will be shown in the rest of this paper, the ERAM detector allows to reach a spatial resolution better than $800~{\rm\text{\textmu} m}$ for all the incident angles and drift distances, and a dE/dx resolution better than 10\% for tracks crossing the entire ERAM module. These performances are in good agreement with the ones predicted by the simulation that is able to reproduce low level variables, such as the charge sharing between neighboring pads or their time difference, as well as the spatial and the dE/dx resolution. 

\vspace{0.2cm}

The paper is organized as follows. 
Section~\ref{sec:ERAMtechnology} introduces the ERAM technology and its use for the HA-TPC prototype.
Section~\ref{sec:setup} presents the experimental setup. The simulation of the ERAM response is described in Section~\ref{sec:simu}. Section~\ref{sec:testbench} is devoted to the characterization of the ERAM detector. The data collected during the DESY test beam are presented in Section~\ref{sec:data}, while Section~\ref{sec:reco} provides a description of reconstruction algorithms. Section~\ref{sec:testbeam} is devoted to a comparison of ERAM response between data and simulation. Spatial and dE/dx resolutions are presented in Sections~\ref{sec:spatial} and~\ref{sec:dedx}, respectively, while Section~\ref{sec:datasimu} discusses a comparison between the data and the simulation. A study of the E$\times$B effect is presented in Section~\ref{sec:ecrossb}. A short discussion on further potential improvements is given in Section~\ref{sec:discussion}. The conclusions in Section~\ref{sec:conclusion} close the paper.  

\section{The ERAM technology and its use for the HA-TPC prototype}
\label{sec:ERAMtechnology}

The ERAM technology, initially developed for the ILC prototypes~\cite{Attie:2011zz}, allows the spreading of the charge induced on the ``leading” pad by the electrons, produced from ionization, over several adjacent ``neighbour" pads. The signal produced by the charge deposited in each pad (or ``waveform") is a function of time and it can be predicted.
The combination of the information from these signals on different pads forming a “cluster” allows us to improve the spatial resolution and hence the determination of the momentum of charged particles. Different ERAM prototypes have been tested at CERN~\cite{Attie:2019hua} and at DESY~\cite{Attie:2021yeh}
to characterize the ERAM response.

\begin{figure}[H]
    \centering
        \includegraphics[width=0.6\linewidth]{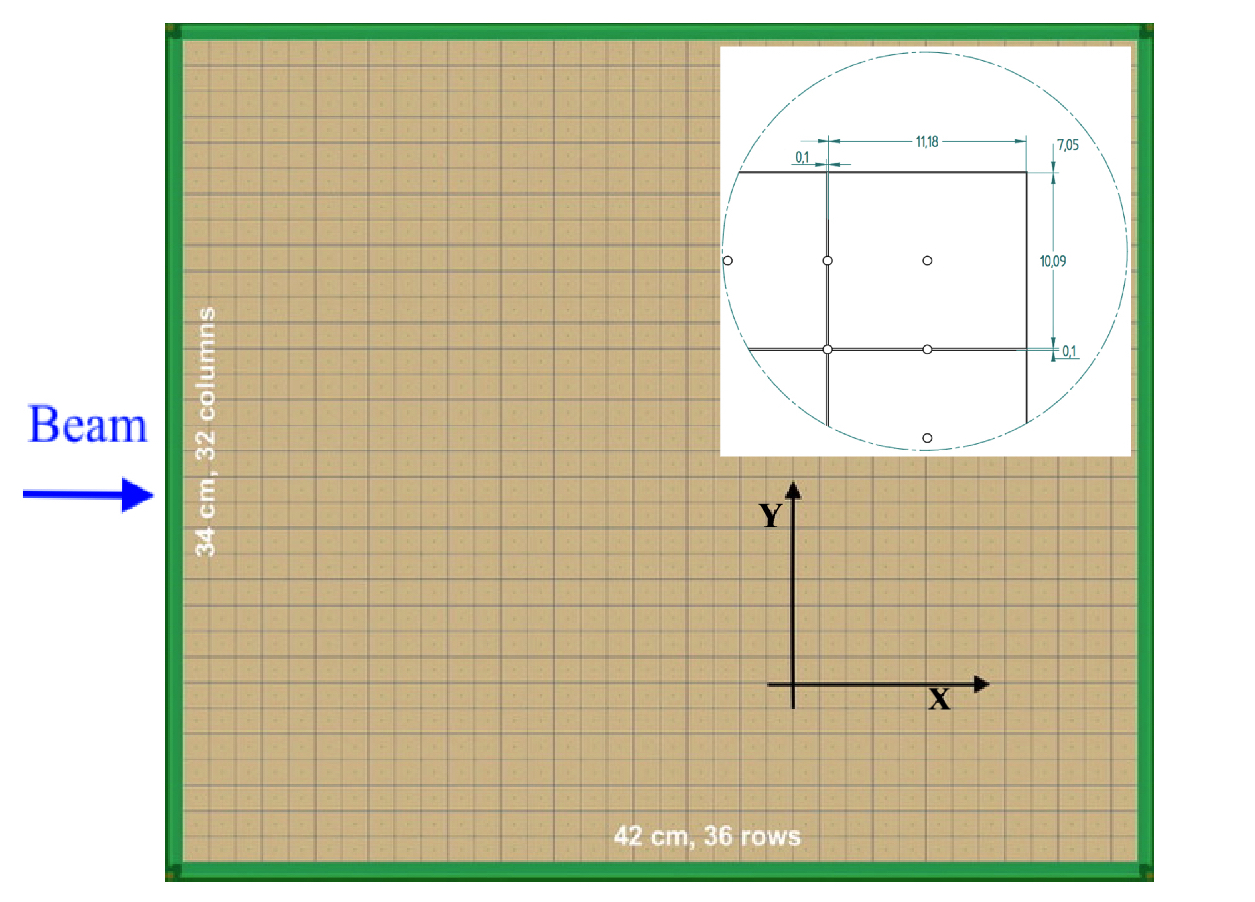} \
    \caption{ERAM technical drawing and its local reference frame.
}
    \label{fig:eram_drawing}
\end{figure}

Throughout this paper we define ``horizontal" and ``vertical" tracks as tracks along the horizontal (X-axis in the local ERAM reference frame) and vertical (Y-axis in the local ERAM reference frame) pad border of ERAM respectively (see Fig.~\ref{fig:eram_drawing}). ``Inclined" tracks are the tracks tilted 
in the X-Y plane.
Thanks to the PCMAG movable stage we could move the TPC prototype, performing scans of the ERAM in Y and X (rotating the prototype by 90$^\circ$) directions as well as scans in Z (drift distance). The global reference frame is provided in Fig.~\ref{fig:orient}. 


In this paper, for track reconstruction we use a method similar to the one already implemented for the analysis of previous test beam periods at CERN~\cite{Attie:2019hua} and at DESY~\cite{Attie:2021yeh}.
This reconstruction method is based on the maximal amplitudes of the waveforms in the pads constituting the clusters. Clusters of pads are built to be transverse to the track projections on the detection plane, following the ERAM columns (rows) for horizontal (vertical) tracks 
or their diagonals for inclined tracks, see 
Figs.~4 and~6 in~\cite{Attie:2021yeh}.  
For each cluster, composed of multiple pads, a track position 
is reconstructed. The spatial resolution refers to the accuracy of the measured position in the cluster
with respect to the reconstructed track position. 
Studies are underway aiming at using simultaneously the full information of the waveforms of all the pads around the tracks in a global fit, to improve the track reconstruction accuracy. A more simple approach used here is a robust reference for further studies,  and can already show that the detector performances meet the physics requirements of the experiment. The results of this test beam campaign allowed us to validate the ERAM design and start the production of the 32 ERAMs that will be used to instrument the HA-TPCs. 

The large drift distance available in the TPC prototype under test, which is similar to the one of the final HA-TPC, is crucial to validate our understanding of two important effects: a ``charge sharing'' between pads induced by the diffusion in the gas and a ``charge spreading'' due to the ERAM resistive foil. Contrary to the ``charge sharing'' effect, the signals in neighbouring pads are delayed compared to the one of the leading pad in case of the ``charge spreading''.

The impact of these effects will be discussed after a detailed presentation of the obtained results.

\label{sec:introduction}

\section{Experimental setup}
\label{sec:setup}

\subsection{HA-TPC field cage prototype}
\label{subsection_fieldC}
One of the main innovations of the HA-TPCs with respect to the vertical TPCs~\cite{T2KND280TPC:2010nnd} currently used in ND280 is that the new field cage will use a single layer of solid insulator laminated on composite material, while for the current ND280 TPCs, two gas-tight boxes, one inside the other, are used. This new design minimizes the dead space and maximizes the tracking volume by reducing the distance between the outer TPC wall and the active gas volume from 12~cm to 4~cm. The radiation length of the material composing the field cage is 2\%.
The used gas composition is the standard T2K gas~\cite{Abgrall:2010hi}, a mixture of Ar:CF$_4$:$\rm iC_{4}H_{10}$ (95:3:2).

In order to test the construction process of the HA-TPCs field cages, several prototypes have been produced. One of them, that shares all the characteristics of the final field cages, was used for the test beam described in this paper.

The field cage prototype is built with lightweight and low-Z mechanical structures with a hollow shell shape constituting the box. The box is laminated on an Aluminum mold in several 
layers, namely Kapton\textsuperscript{\textregistered} sheets, aramide fiber-fabrics peels and honeycomb spacer panels
glued together. 
The field cage is then enclosed on two sides with a cathode plane and the anode where the ERAM detector is located.

The innermost cage wall surface embeds a double layer of thin copper strips: 
the field strips for degrading the potential from the cathode to the anode
and the mirror strips on the opposite side, for regularizing the field nearby the walls 
and for mitigating the effects of free charge deposition on dielectric surfaces.
The strip foils are produced by the CERN Micro-Pattern Technologies service. In order to protect the field cage from the possible presence of tiny carbon fibers 
embedded into the aramide fiber fabric (Twaron)  the mirror strip side is protected with an additional Kapton\textsuperscript{\textregistered} coverlay glued on it. 

\begin{figure}[!ht]
    \centering
    \begin{minipage}{0.98\linewidth}
        \centering
        \includegraphics[width=\linewidth]{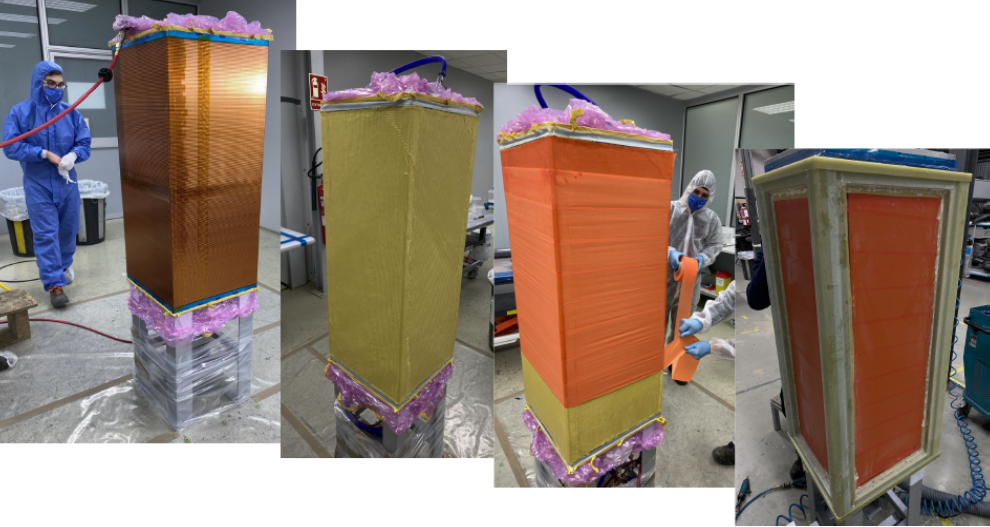} 
    \end{minipage}
    \caption{Phases of a field cage prototype production. From left to right: strip foil wrapped onto the mold, aramid fiber fabric (Twaron) glued onto the strip foil layer, Kapton\textsuperscript{\textregistered} tape wrapped on the Twaron layer before glue curing phase, top and bottom flanges and angular bars applied. The vertical direction in these photos represents the drift direction and the ERAM will be installed on the top surface with the cathode on the bottom.}
    \label{fig:fc_prototype_production}
\end{figure}

The prototype has the same drift length ($97.25$~cm) 
as the HA-TPCs and a reduced transverse area ($42\times 42$~cm$^2$) suitable to host one ERAM module. 
It was produced by the NEXUS company (Barcelona, Spain) and the different phases of the production of the prototype are shown in Fig.~\ref{fig:fc_prototype_production}.

Prior to exposing the prototype to the electron beam at DESY, an extensive characterization of the field cage has been done at CERN. The properties were extremely good concerning fiberglass flanges smoothness quality, 
gas tightness (measured leakage below 0.1 l/h), inner surface quality and deformations
smaller than 0.2~mm,  compatible with prototype mold tolerances. We also performed several measurements of resistance and capacitance on the field and mirror strips showing that the electric field is behaving as expected.

\subsection{ERAM detector}

A description of the ERAM technology and of the detector used for the HA-TPCs of T2K is given in~\cite{Attie:2021yeh}. 
The ERAM modules built for T2K have a size of $42 \times 34$ cm$^2$ and are segmented in $36\times32$ rectangular pads of size $11.18 \times 10.09$~mm$^2$. 

The ERAMs are used to readout the ionization electrons produced by charged particles crossing the TPC gas volume. These electrons are drifted to the anode readout plane of the TPC under a uniform electric field. On the readout plane, an avalanche is generated by a high electric field in the ERAM amplification region. The resulting pattern of illuminated pads corresponds to the trajectory of the track.

The main difference between the bulk-Micromegas technology used for the existing ND280 TPCs and an ERAM is that, in the case of bulk-Micromegas and for short drift distances the position reconstruction is limited by the pad size that is large compared to the size of the avalanche which falls on a metallic anode.  In the ERAM, instead, the anode is covered by a foil of insulating material with a thin resistive layer on top, inducing signals over several pads. This allows a better reconstruction of the position of the charged particles crossing the TPC.

The ERAM detector uses a Diamond-Like Carbon (DLC) thin layer sputtered on a 50 ${\rm\text{\textmu} m}$ thick APICAL\textsuperscript{\textregistered} (Kapton\textsuperscript{\textregistered}) insulator sheet.
The detector installed on the field cage, named ERAM-01, has a resistivity of 300--400 kOhm/$\Box$ using DLC foils stack on a 150 ${\rm\text{\textmu} m}$ glue layer.

\subsection{HA-TPC electronics}

The full electronics chain that will be used for the HA-TPCs has been installed on the ERAM-01 and tested during the test beam described in this paper. One of the goals of the test beam campaign was to validate the 
HA-TPC front-end electronics 
performance when placed inside a 0.2 T magnetic field. It has been then proved that their behavior is not altered. 

\begin{figure}[!tpb]
\begin{center}
\includegraphics[width=0.3\linewidth]{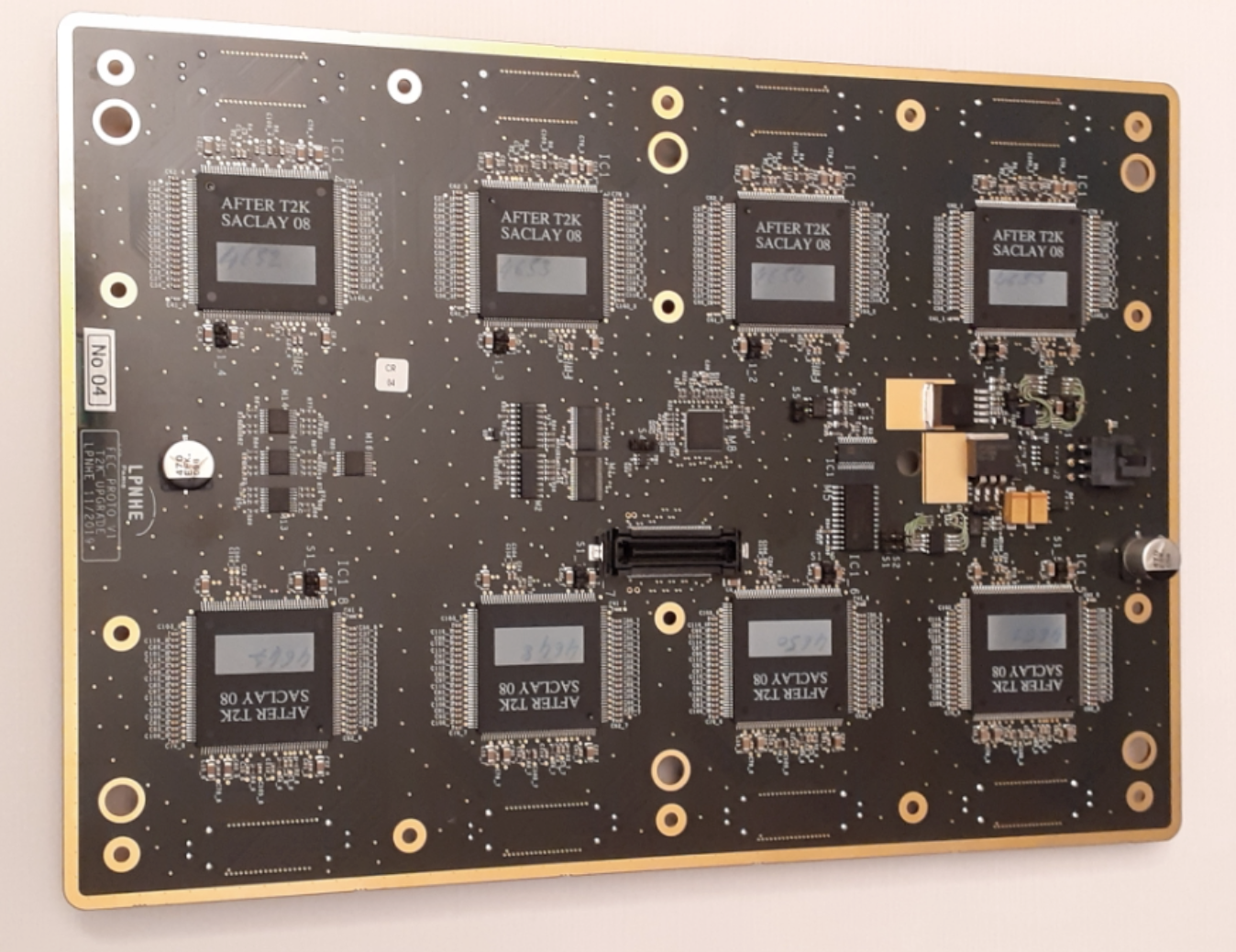}\hfill
\includegraphics[width=0.3\linewidth]{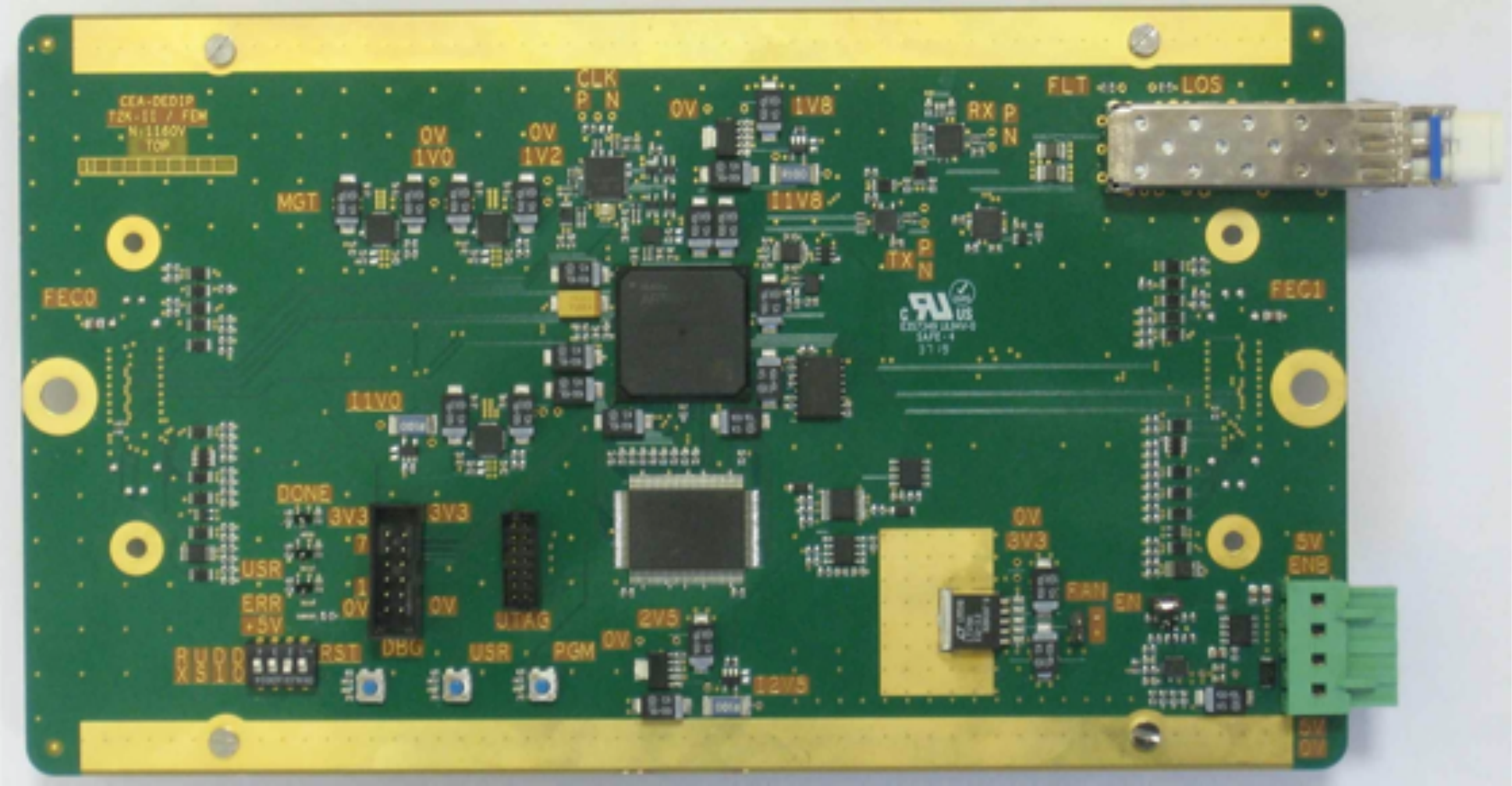}\hfill
\includegraphics[width=0.3\linewidth]{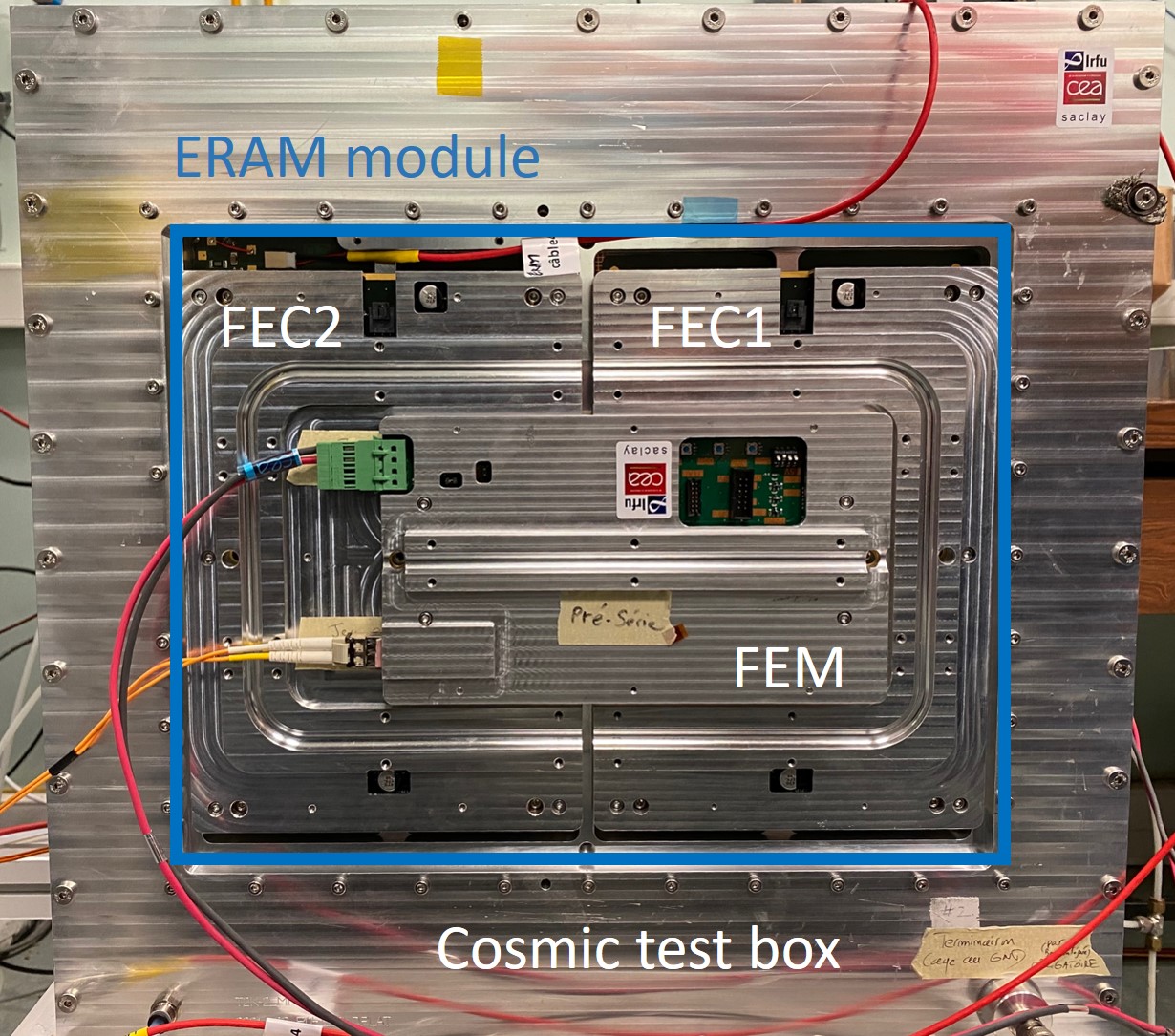}

\caption{Photo of one FEC with the 8 AFTER chips (left), and of one FEM (center). The right picture illustrates the final assembly of the ERAM with two FEC boards connected to one FEM; 
all electronic boards are equipped with their cooling plates.}\label{Fig:fecfem}
\end{center}
\end{figure}

The HA-TPC electronics, shown in Fig.~\ref{Fig:fecfem}, is based on the use of the AFTER chips~\cite{Baron:2008zza}, that had been designed for the existing ND280 vertical TPCs. 
The AFTER chip is a 72-channel ASIC that includes preamplifiers and shapers with programmable gain and peaking time coupled to a 511-time bucket switched capacitor array (SCA). During the test beam the electronics peaking time was set to either 200 or 412~ns. 

The FECs have been newly designed and host 8 AFTER chips. They are installed parallel to the ERAM modules and two FECs are used to readout one ERAM (1152 channels). The response linearity of the FEC has been measured with a dedicated campaign and showed a uniform response of all the channels with typical differences in linearity among neighboring pads smaller than 2\%.
The two FECs on each ERAM are connected to a FEM card that performs their control, synchronization and data aggregation.
All connections, firstly between ERAM and FEC boards, 
secondly between FEC and FEM boards, are performed by using ``floating'' 
type connectors (HIROSE - FX23/FX23L series) in order to eliminate wired connections and all their drawbacks.
The final production of electronic boards - both FECs and FEMs - 
was performed by the OUESTRONIC company in Rennes, France~\cite{OUESTRONIC}.

As back-end electronics we used the TDCM, a generic clock, trigger distributor and data aggregator module~\cite{Calvet:2018lac} designed for several projects, including the HA-TPCs.
The collected data are transferred to a DAQ computer via a MIDAS~\cite{MIDAS} front-end and stored on disk for further analysis.

\section{Simulation of the ERAM response}
\label{sec:simu}
In this paper we  present the first comparisons between the data and a Monte Carlo simulation of the ERAM detector. To develop the simulation we benefited from the already extensively verified simulation of the vertical TPCs in the ND280 software~\cite{Abe:2011ks}. The new feature that was implemented in the model for this work is the resistive layer of the ERAM detectors.

\subsection{Simulation framework}
The simulation starts from GEANT4~\cite{GEANT4:2002zbu} that is used to propagate the charged particles in the TPC gas and to produce the ionization electrons (typically 100 electrons per cm). To simulate the fluctuations in the ionization the PAI model~\cite{Apostolakis2000} is used with at most a 1 mm computation step.

These electrons are individually transferred to the ERAM plane following a straight line. The position and arrival time of each ionization electron are distributed on the sensitive plane following Gaussian distributions assuming a transverse diffusion of $\sigma_{trans}=286\ {\rm\text{\textmu} m} / \sqrt{\rm{cm} }$ and a longitudinal diffusion of $\sigma_{long}=210\ {\rm\text{\textmu} m} / \sqrt{\rm{cm}}$. For each electron arriving to the ERAM the amplification is simulated based on the ERAM gain $\rm G$. Fluctuations in the avalanche processes are taken into account by extracting the gain $\rm g_e$ for each electron as $\rm{g_e} =- \rm{log}\left(1- \rm{uniform}(0, 1)\right) \times G$, hence assuming an exponential gain. 

The amplified signal is then given as input to the simulation of the resistive layer that will be introduced in the next section. The resulting signal in each pad is then convoluted with the AFTER chip electronics response function and digitized with a sampling time of 40 ns. 

\subsection{Resistive layer simulation}
\label{sec:rc_sim}

The behavior of the resistive layer can be approximated to a RC continuous network~\cite{Dixit:2006ge}. In this model, the charge density caused by the point-like electron deposited in $ \Vec{\rm r_0} = (\rm x_0, y_0)$ $(\rho(\Vec{\rm r}, {\rm t=0}) = \delta_{\Vec{r_0}}(\Vec{\rm r}) )$ is described with the solution of the 2D diffusion equation:

\begin{equation}
    \rho(\Vec{r},t)=\frac{RC}{4 \pi t} \times\exp \left(-\frac{r^2 RC}{4t} \right) 
\end{equation}
where $ \rm r = \sqrt{(x-x_0)^2+(y-y_0)^2}$ is a distance from the initial charge deposition, $\rm t$ is time and $\rm RC$ is a network characteristic of the ERAM. For our case, ERAM is expected to have an RC within 50-120 $\rm ns/mm^2$. 
To compute the observed charge in a given pad the equation above should be integrated over the pad surface

\begin{equation}
\begin{split}
    Q_{unit}(t) = & \int _{x_{min}}^{x_{max}}  \int _{y_{min}}^{y_{max}}  \rho(\Vec{r},t) \mathrm{d}x~\mathrm{d}y  \\
     = & \frac{1}{2}\pi\left(Erf\left[\frac{\sqrt{RC}(x_{max}-x_0)}{2\sqrt{t}}\right]
    -Erf\left[\frac{\sqrt{RC}(x_{min}-x_0)}{2\sqrt{t}}\right]\right) \\
    & \times \left(Erf\left[\frac{\sqrt{RC}(y_{max}-y_0)}{2\sqrt{t}}\right]-Erf\left[\frac{\sqrt{RC}(y_{min}-y_0)}{2\sqrt{t}}\right]\right)
\label{eq:qresp}
\end{split}
\end{equation}
where $\rm Erf$ is the error function and $\rm x_{min},\ x_{max},\ y_{min},\ y_{max}$ are pad borders coordinates.

The evolution of the charge in the pad is convoluted with the derivative of the AFTER electronics response: 

\begin{equation}
    \label{eq:eresp}
    E(t)=\left(\frac{t}{t_{p}}\right)^3\exp\left(-\frac{3t}{t_{p}}\right)\sin\left(\frac{t}{t_{p}}\right)
\end{equation}

where $\rm t_p$ is the electronics peaking time. The unit waveform ($\rm WF_{unit}$) is then:

\begin{equation}
\label{eq:eram_resp}
    WF_{unit}(t)=Q_{unit}(t)\circledast \frac{dE}{dt}(t) = \int _{-\infty} ^{\infty}  Q_{unit}(t-\tau)  \frac{dE}{dt}(\tau) \mathrm{d}\tau.
\end{equation}

When considering an avalanche of electrons, the waveforms induced in each pad by each electron of the avalanche should be computed taking into account the arrival time of each of them and summed in order to obtain the complete waveform ($\rm WF$) of the avalanche.

Therefore, the numerical evaluation of the diffusion equation solution and of the convolution is extremely heavy in terms of computation time. To keep the simulation to a reasonable time some approximations were included as described below. 

The first method is related to reducing the total number of avalanches to be simulated. The pad is divided into several smaller sub-pad regions e.g. 3$\times$3 or 5$\times$5. All the avalanches that are detected in the same sub-pad are merged into one and the charge Q is computed for the sum of all the contributions in this sub-pad. 

The next and most significant optimization is related to the pre-computation of the diffusion equation solution and convolution (as in Eq.~\ref{eq:eram_resp}). 
Before starting the simulation the detector response is pre-computed for a unit charge, a given RC that is input to the model, and for all the positions across a 2D grid in the pad, $\rm Q_{unit}(RC, x_i, y_i, t)$, where $\rm x_i$ and $\rm y_i$ are the coordinates of a sub-pad center. 
The step of the grid can be tuned and for this work we divided the pad in a grid of 10$\times$10 sub-pad regions. 
The obtained distributions are convoluted with the derivative of the electronics response to get the waveform ($\rm WF_{unit}$) for a unit charge. 
The final waveform $\rm WF(t)$ can be easily obtained by scaling the pre-computed solution $ \rm WF_{unit}$ with the total charge $\rm Q_i$ in each sub-pad $\rm i$, so that:

\begin{equation}
WF(t)=\sum_i Q_i\times WF_{unit}(x_i, y_i)(t).
\end{equation}

The main approximation in this computation is the assumption that all the electrons in a sub-pad arrive at the center of this sub-pad.
This implies that no numerical computations are needed during the simulation. 

The described optimization methods reduce the required CPU time by more than two orders of magnitude without noticeable impacts on the simulation output.

\begin{figure}[t]
    \centering
    \begin{minipage}{0.49\linewidth}
        \centering
        \includegraphics[width=\linewidth]{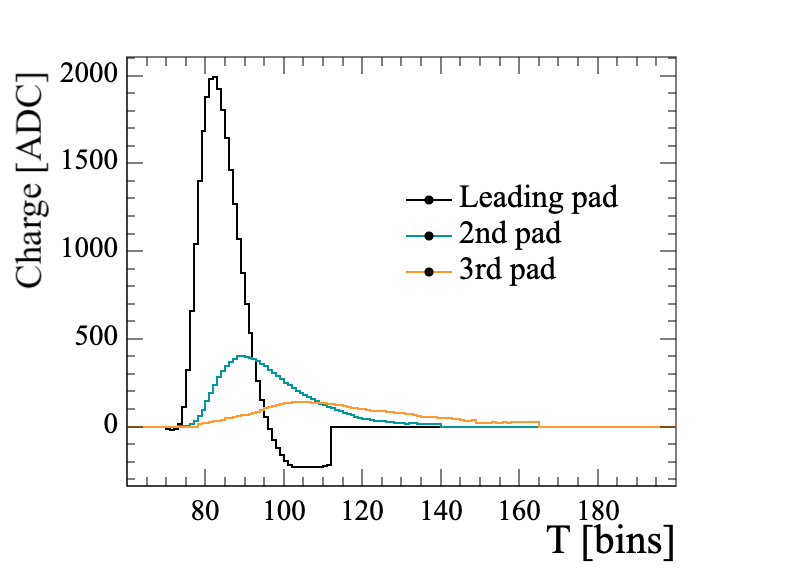} \\ a
    \end{minipage}
    \begin{minipage}{0.49\linewidth}
        \centering
        \includegraphics[width=\linewidth]{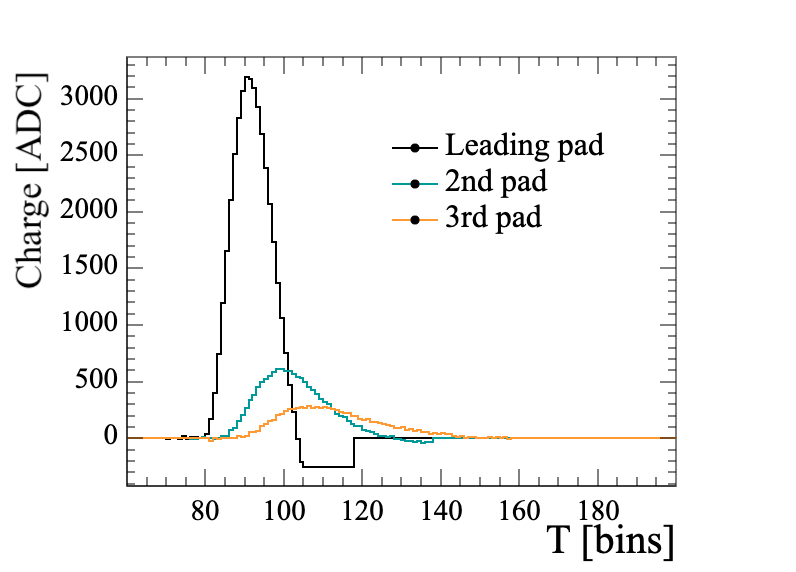} \\ b
    \end{minipage}
    \caption{The example waveforms for the leading and adjacent pads in (a) data and (b) MC. The truncation of the underflow is due to the fact that in both, data and simulations, pedestals for each channel are equalized to 250 and, when the waveform goes below 0 they are set to 0. The 250 baseline is then subtracted offline and is not shown in the plot.}
    \label{fig:mc_wf}
\end{figure}

Two examples of the WF, one for the data and one for the simulation, are shown in Fig.~\ref{fig:mc_wf} for the leading and for two neighboring pads. Here and in the following of this paper, the leading pad is defined as the pad with the largest maximum of the waveform while neighboring pads are the ones adjacent to the leading pad in the direction perpendicular to the track projection on the ERAM plane (see discussion on clustering algorithms in Sect.~\ref{sec:reco}).

\section{Characterization of ERAM detector}
\label{sec:testbench}

Each ERAM detector is scanned after production using an X-ray test bench at CERN. The test bench consists of a 3~cm wide gas chamber and a robotic X-Y-Z arm system on an optical breadboard of 120$\times$60~cm$^2$ holding a 250~MBq $^{55}\rm Fe$ source emitting 5.9~keV photons that deposit all their energy in a gas volume filled with the standard T2K gas mixture.

A 1.5~mm diameter collimation hole in front of the source assures that the majority of photo-electrons arrive on the targeted pad. Prior to the scan, the X-ray source is aligned to ensure that it is placed in front of the center of each ERAM pad.

\begin{figure}[H]
    \centering
    \centering
    \begin{minipage}{0.49\linewidth}
        \centering
        \includegraphics[width=0.82\linewidth]{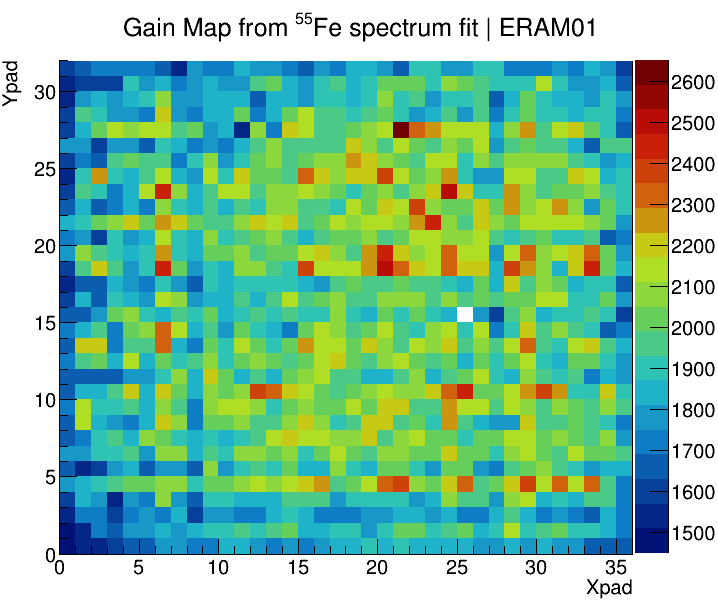}\\ a
    \end{minipage}
    \begin{minipage}{0.49\linewidth}
        \centering
        \includegraphics[width=1\linewidth]{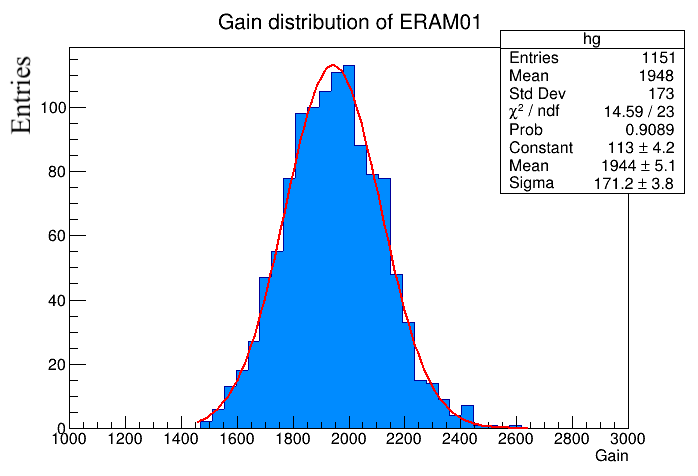} \\ b
    \end{minipage}
    \caption{
        ERAM-01 XY-map of the gain (a) and gain distribution (b) obtained with a $^{55}\rm Fe$ source. The channel (25,16) is a dead channel.
    }
    \label{fig:testbench_results}
\end{figure}

Each channel of the ERAM is scanned for $\sim$3 minutes at a rate of 100~Hz, to reconstruct the spectrum of the $^{55} \rm Fe$ source and compute the gain for each pad. The gain is computed as the mean of the sum of the waveforms in a 3$\times$3 matrix around each channel. The map of the gain on the ERAM-01 and the gain uniformity are shown in Fig.~\ref{fig:testbench_results}. 

The data from the test bench can also be used to measure the RC uniformity of the ERAM. This measurement campaign is a subject of a dedicated publication~\cite{Attie:2023uqx}. In simulation we assumed uniform gain and RC.

\section{Data collected at DESY}
\label{sec:data}
The tests at DESY were aimed to ensure that the HA-TPC prototype design fully satisfies the requirement of the ND280 upgrade for drift distances up to 97.25~cm that corresponds to the maximum drift length of the HA-TPC. Therefore, this test beam covered all the possible tracking conditions of HA-TPC.

The HA-TPC prototype equipped with ERAM-01 was flushed with T2K gas mixture and tested at DESY T24/1 facility. The chamber was placed inside the PCMAG solenoid (Fig.\ref{fig:orient}) providing a magnetic field up to 1.25~T and exposed to the electron beam with tunable momenta between 1 and 4~GeV/$c$. The solenoid is equipped with a movable stage that allows moving the detector along the horizontal and vertical directions.

Most of the data were taken with the cathode high voltage set at 26.7~kV corresponding to an electric field in the TPC of 275 V/cm and at $\rm B=0.2$~T as used in ND280. This field configuration corresponds to a drift velocity of 7.9 $\rm cm/{\rm\text{\textmu}} s$.
In order to study the dependence on different parameters and configurations, various scans were performed. 

X and Y scans are particularly interesting because they allow to study the impact on the performances of possible non-uniformities in the gain or in RC.

The drift distance scan was done for two values of the electronics peaking time of 412 ns and 200 ns. 

The data were also collected for different rotation angles around the Z-axis ($\phi$ angle) and for three drift distances corresponding to three different points of origin of the ionization, one close to the ERAM, one in the middle of the chamber, and one close to the cathode.

\section{Reconstruction and selection of tracks}
\label{sec:reco}
Tracks in simulated events and real data were reconstructed with the same analysis framework that uses Density-Based Spatial Clustering of Applications with Noise (DBSCAN)~\cite{Ester96adensity-based} algorithm and the “pad response function” (PRF) method (discussed in Sect.~\ref{sec:spatial}). 

The track projection on the ERAM plane is composed by clusters that are groups of pads in the direction perpendicular to the track projection.
In order to be selected, a track needs to cross the whole detector.
Due to a large number of pads per cluster (multiplicity) induced by the resisitive layer, two close parallel tracks may not be separated by a gap and thus can be mis-reconstructed as one single track. 

To reject such a topology and also to remove superimposed tracks, a cut on the mean multiplicity of the track is applied. This cut depends on the track clustering algorithm that is used to reconstruct the tracks. As it was introduced in~\cite{Attie:2021yeh}, we use horizontal and vertical clustering for tracks with the angles of inclination w.r.t. the ERAM plane below 30 degrees and above 60 degrees respectively, while for inclined tracks we use a diagonal clustering algorithm in which pads are combined into clusters according to their diagonal.

\begin{figure}[!ht]
    \centering
    \begin{minipage}{0.49\linewidth}
        \centering
        \includegraphics[width=\linewidth]{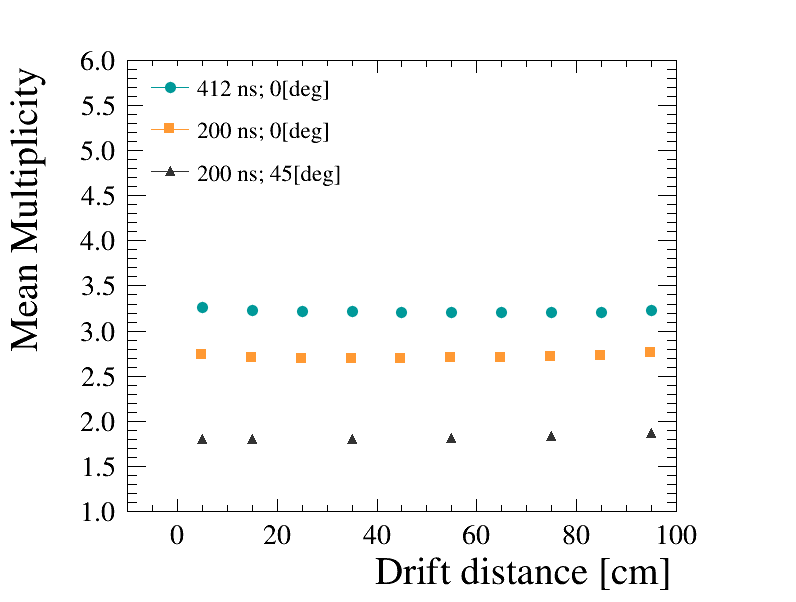} \\ a
    \end{minipage}
    \begin{minipage}{0.49\linewidth}
        \centering
        \includegraphics[width=\linewidth]{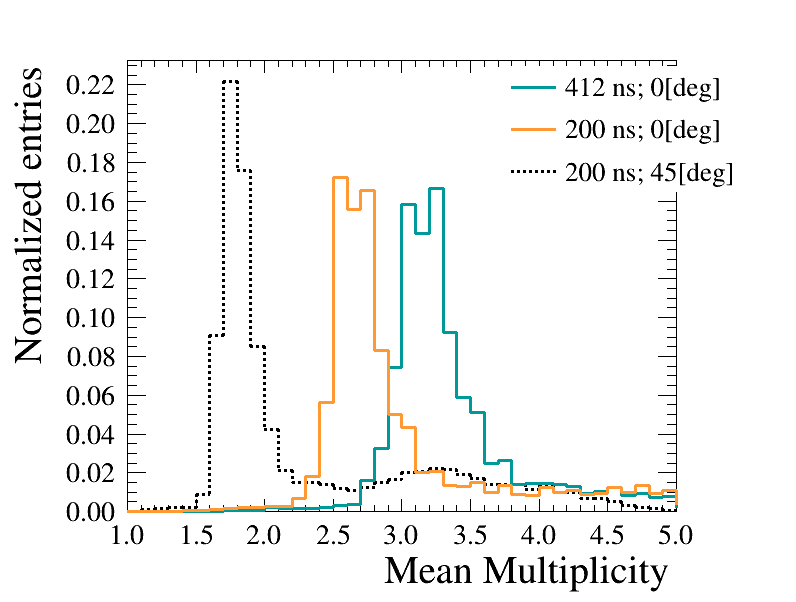} \\ b
    \end{minipage}
    \caption{Mean multiplicity as a function of drift distance (a) and mean multiplicity (b) at different peaking times for horizontal and inclined tracks.}
    \label{fig:mean_multi}
\end{figure}

The mean multiplicity depends on the reconstruction algorithm and on the peaking time for the electronics. It is shown in Fig.~\ref{fig:mean_multi} for horizontal and inclined tracks at various drift distances and peaking times. 

For the analyses presented in this paper, we select tracks with mean multiplicity comprised between 2.4 and 3.2 (1.5 and 2.2) for horizontal (inclined) tracks with 200~ns peaking time. For 412~ns peaking time only horizontal tracks were taken and, for this sample, we required a mean multiplicity comprised between 2.7 and 3.6. 

Finally, to avoid edge effects, the pads at the border are excluded from the reconstruction and hence horizontal tracks have 34 clusters while vertical tracks have 30.

\section{ERAM response in data and simulation}
\label{sec:testbeam}

To validate the simulation described in Sect.~\ref{sec:simu}, we produced electrons crossing the HA-TPC prototype and compared some low-level variables describing the ERAM response, including charge sharing and time differences between neighboring pads in data and simulation. 
The comparison between data and simulation for spatial resolution and dE/dx resolution will be shown in Sect.~\ref{sec:datasimu}.

The most important variables for the characterization of resistive feature are the charge ratio and the time difference between the signals observed in the adjacent pads and the one in the leading pad. 

For these comparisons, the charge in the pad is defined as the maximum of the waveform and the time is defined as the time bin at which the waveform reaches its maximum. The pads in a cluster are then ordered according to their charge ($\rm Q_1$ and $\rm T_1$ refer to the pad with the largest charge, $\rm Q_2$ and $\rm T_2$ to the second pad, and so on).

\begin{figure}[!ht]
    \centering
    \includegraphics[width=0.32\linewidth]{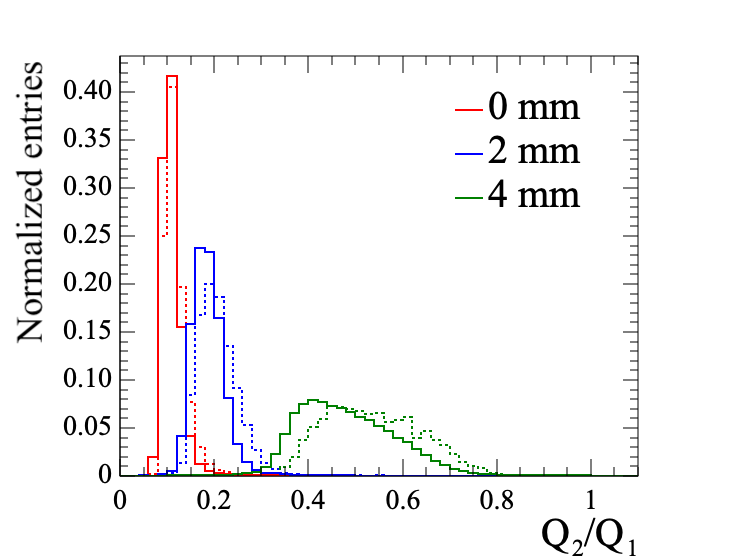}
 \includegraphics[width=0.32\linewidth]{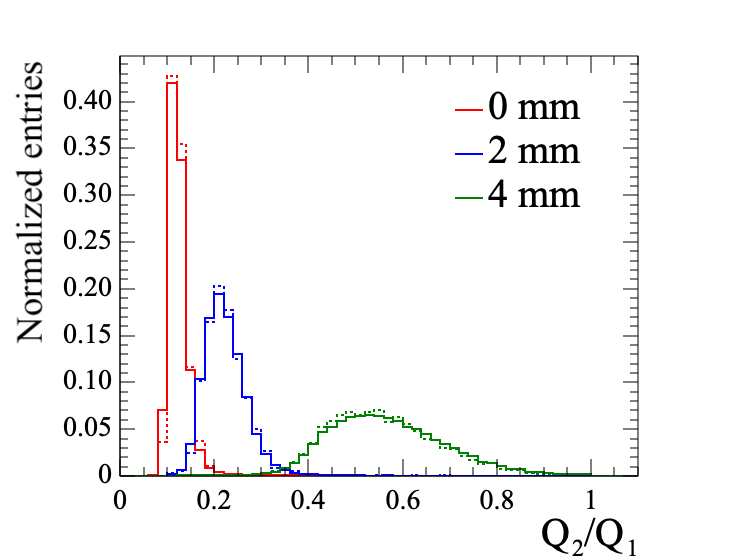}
  \includegraphics[width=0.32\linewidth]{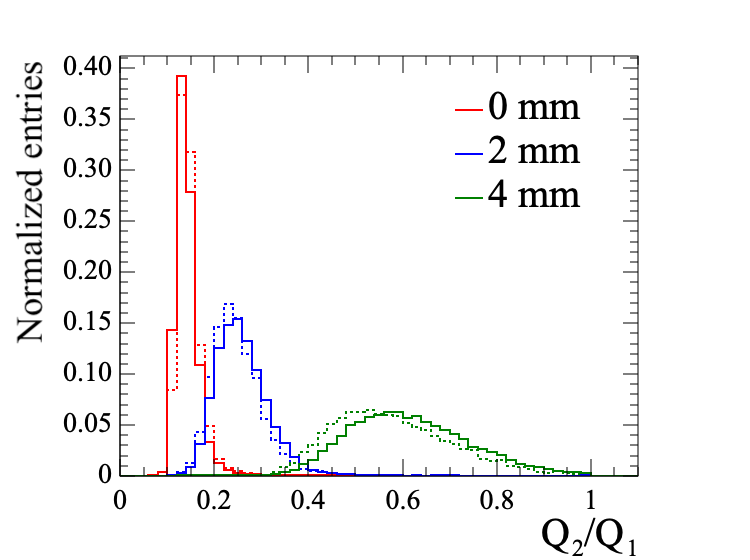}
    \includegraphics[width=0.32\linewidth]{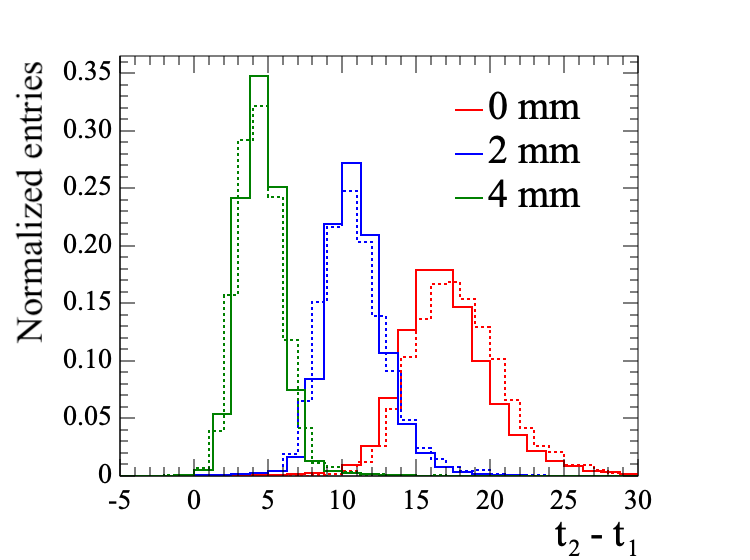}
  \includegraphics[width=0.32\linewidth]{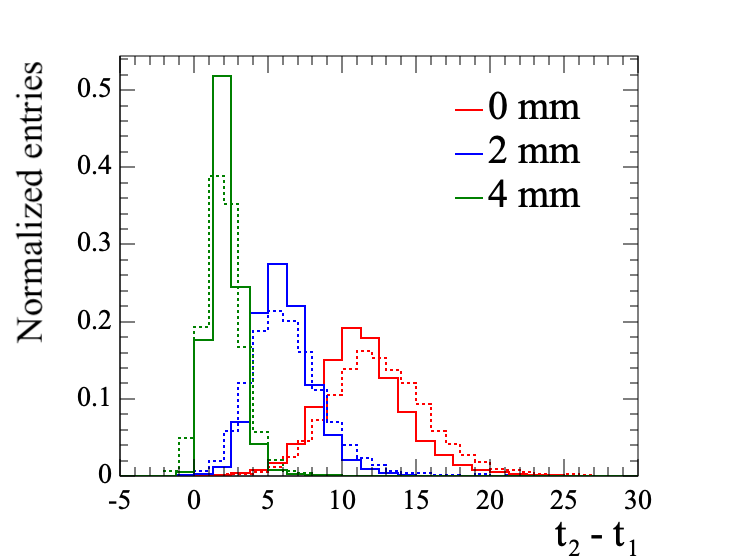}
  \includegraphics[width=0.32\linewidth]{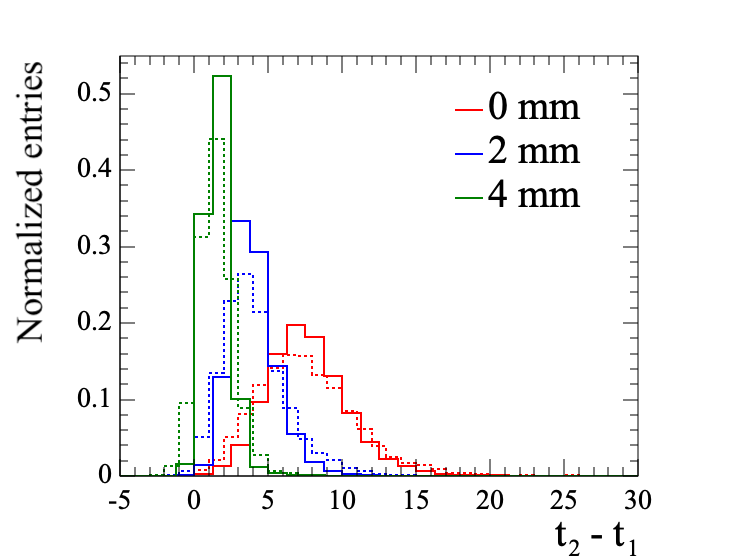}

  \caption{(Top) ratio of the charge in the second pad with respect to the leading pad. (Bottom) time difference between leading and second pads. The distributions are obtained for different reconstructed distances from the pad center. Left plots correspond to the drift distances of 5 cm, central plots of 50 cm and right plots of 90 cm. The solid lines represent data and the dashed ones are the simulation.}
    \label{fig:mc_data_5cm_distr}
\end{figure}

Since these variables depend on the relative position of a track with respect to the pad center, we grouped the distributions based on the reconstructed track position. These distributions are shown in Fig.~\ref{fig:mc_data_5cm_distr} for different drift distances. 

For future analysis, we expect to improve the agreement between data and MC by using in the simulation the value of RC that will be measured with the Test Bench described in Sect.~\ref{sec:testbench} and taking into account possible non-uniformities within the detector.

\section{Spatial resolution}
\label{sec:spatial}

\renewcommand{\degree}{${}^\circ$}

The ND280 TPCs measure the momenta of outgoing particles from neutrino interactions in order to reconstruct the energy of the incoming neutrino, one of the critical elements for precise measurement of neutrino oscillations parameters.

The TPC momentum resolution depends on the spatial resolution~\cite{GLUCKSTERN1963381} that can be characterized with test beam data.
For this analysis, the spatial resolution is determined by employing a PRF method in the same manner as in~\cite{Attie:2019hua,Attie:2021yeh}. 

In this method we define, for each cluster, the residual as the difference between the position of the track reconstructed locally (i.e. in one cluster) and the fitted track position. The distribution of the residuals in each cluster is fitted with a Gaussian and its width represents the spatial resolution. 

The measurement of the track position is performed with an iterative procedure. For the first step, track position in the clusters is reconstructed using the charge barycentric method. Such a method estimates the position of the track in a certain cluster by weighting the centre of the pad position by the charge in this pad. The estimated primary track positions in each cluster are then fit with a parabola over the whole detector (global fit). Based on the results of the fit, a pad response function scatter plot is filled for each pad. The PRF function is defined as:

\begin{equation}
   PRF(x_{track} - x_{pad}) = Q_{pad}/Q_{cluster}
\end{equation}

where $\rm x_{track}$ and $\rm x_{pad}$ are positions of the track from the global fit and the center of the pad, respectively, and $\rm Q_{pad}$ and $\rm Q_{cluster}$ are charges collected by the pad and by the whole cluster. 

The PRF scatter plot is fitted with a ratio of two polynomials (the same as in~\cite{Attie:2021yeh, Boudjemline:2006hf}). The scatter plot and the parametrization of the PRF are done independently for samples at different drift distances and inclinations. The estimated parameters of the PRF analytical function are used further in the $\chi^2$ minimization procedure to estimate the track position in each cluster with: 

\begin{equation}
    \chi^2 = \sum_{pads} \left(\frac{Q_{pad}/Q_{cluster} - PRF(x_{track}-x_{pad})}{\sigma_{Q_{pad}/Q_{cluster}}}\right)^2
\end{equation}

where $\sigma_{\rm Q_{pad}/Q_{cluster}} = \sqrt{\rm Q_{pad}/Q_{cluster}}$.\par
\begin{figure}[!ht]
    \centering
    \begin{minipage}{0.49\linewidth}
        \centering
        \includegraphics[width=\linewidth]{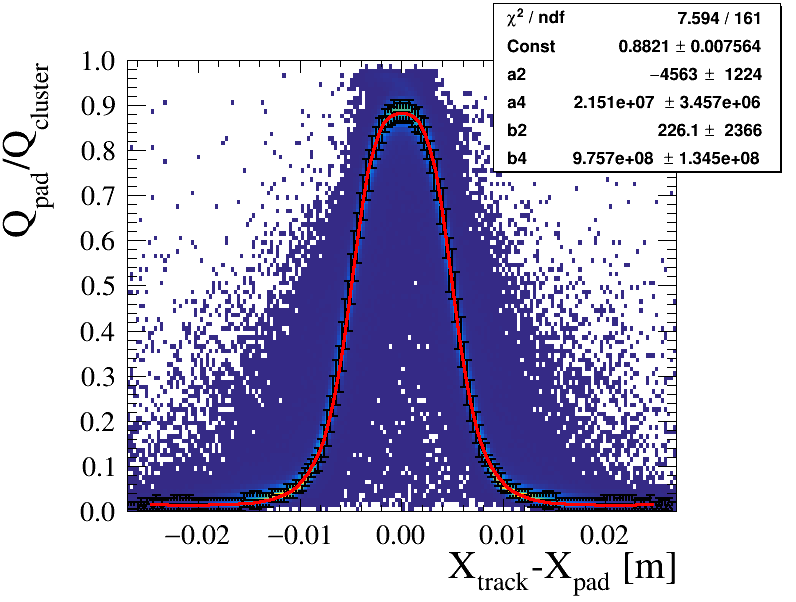} \\ a
    \end{minipage}
    \begin{minipage}{0.49\linewidth}
        \centering
        \includegraphics[width=\linewidth]{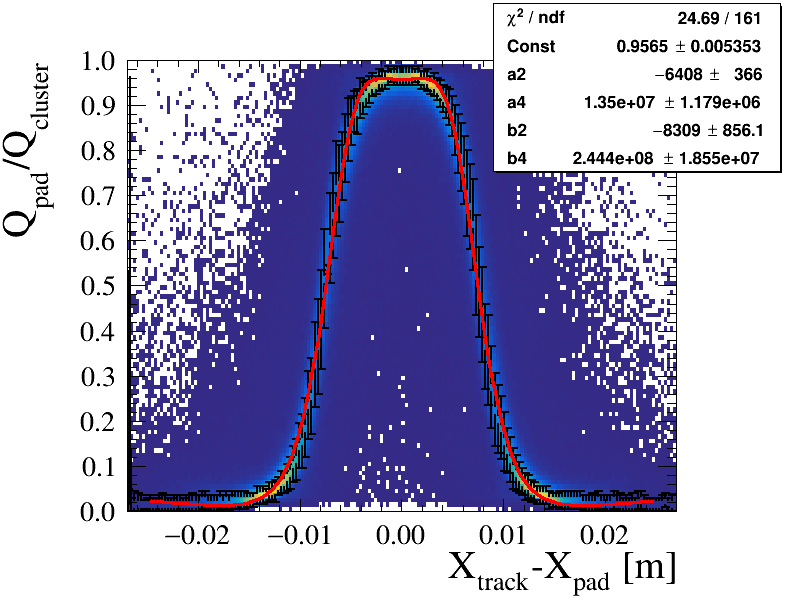} \\ b
    \end{minipage}
    \caption{PRF function for (a) horizontal and (b) inclined tracks.}
    \label{fig:PRF_func}
\end{figure}

In the following iterations the track position is evaluated from the fit. The iterative procedure is repeated while the spatial resolution keeps improving and it typically converges after three iterations. Examples of PRF  for horizontal and inclined tracks are shown in Fig.~\ref{fig:PRF_func}. 

\begin{figure}[!ht]
    \centering
    \begin{minipage}{0.49\linewidth}
        \centering
        \includegraphics[width=\linewidth]{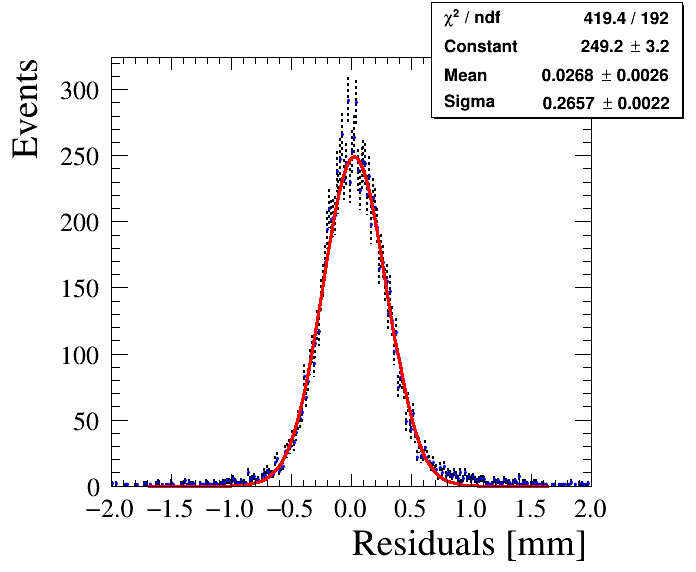} \\ a
    \end{minipage}
    \begin{minipage}{0.49\linewidth}
        \centering
        \includegraphics[width=\linewidth]{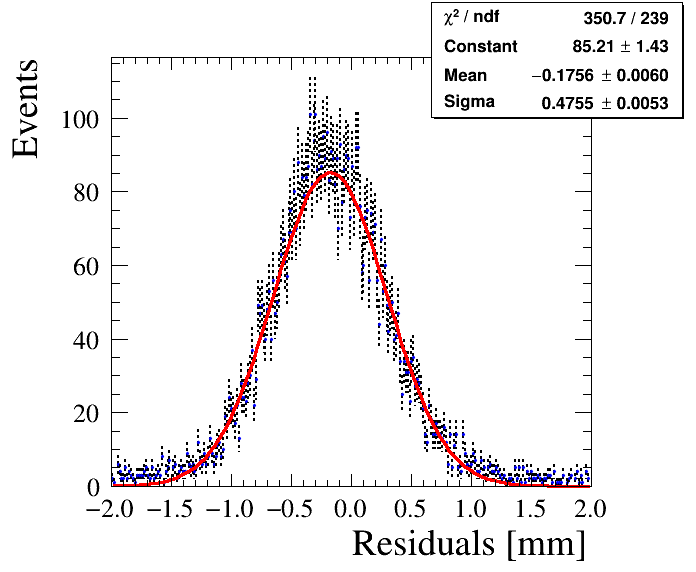} \\ b
    \end{minipage}
    \caption{Residuals distribution for single cluster for (a) horizontal and (b) inclined tracks.}
    \label{fig:PRF_resolution}
\end{figure}

The spatial resolution is defined as the width of the residual distribution for each cluster. Examples for horizontal and inclined tracks are shown in Fig.~\ref{fig:PRF_resolution}. The distribution is expected to be centered at zero and differences with respect to zero are the biases that will be discussed in Sect.~\ref{sec:sr_biases}.

\begin{figure}[!ht]
    \centering
    \begin{minipage}{0.49\linewidth}
        \centering
        \includegraphics[width=\linewidth]{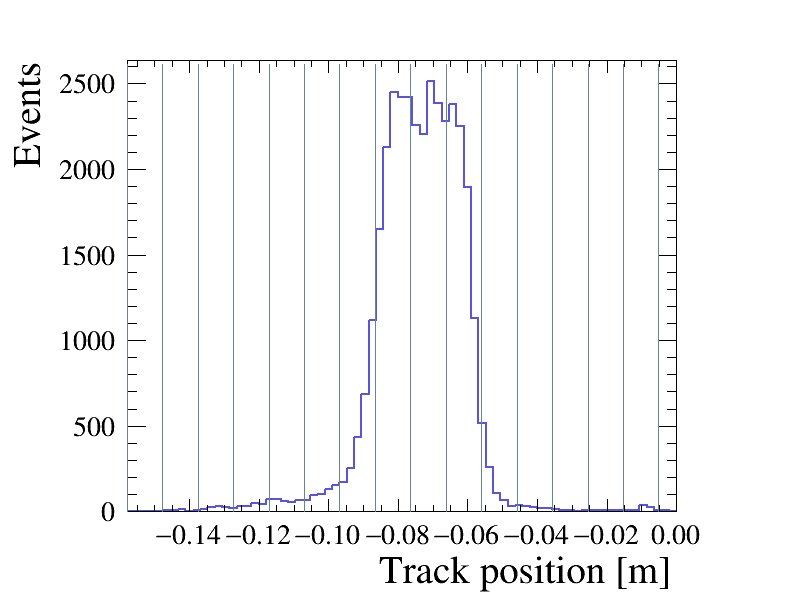} \\ a
    \end{minipage}
    \begin{minipage}{0.49\linewidth}
        \centering
        \includegraphics[width=\linewidth]{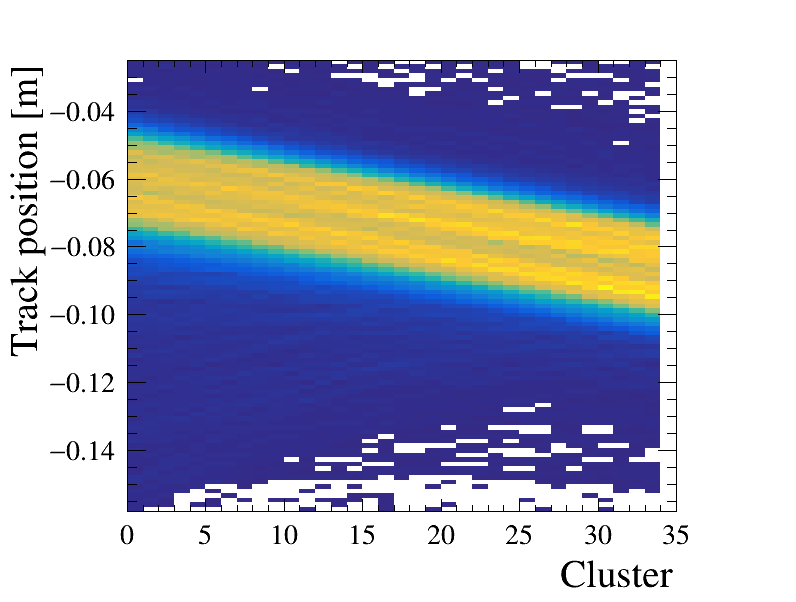} \\ b
    \end{minipage}
    \caption{A typical distribution of the reconstructed track position for incoming electron beam (a), where the vertical lines correspond to the center of each pad and reconstructed track position with respect to the cluster number (b).}
    \label{fig:track_profile}
\end{figure}

As an external cross-check of the performance of tracking algorithm, typical distributions of reconstructed track positions for incoming electron beam are shown in Fig.~\ref{fig:track_profile}. 

\subsection{Spatial resolution for horizontal tracks}
\begin{figure}[H]
    \centering
        \includegraphics[width=0.6\linewidth]{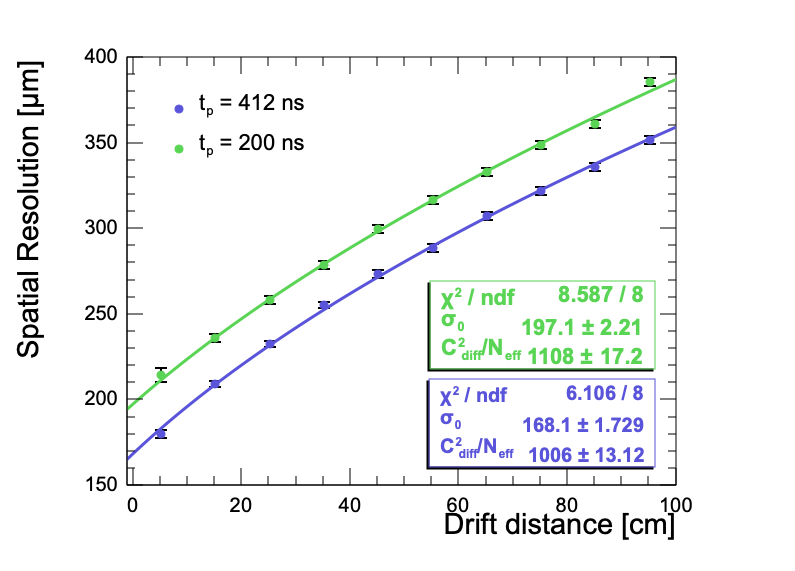} \
    \caption{
        Spatial resolution with respect to the drift distance for horizontal tracks with a magnetic field of 0.2 T, and peaking times of 200 ns and 412 ns.
}
    \label{fig:SR_vs_Z}
\end{figure}

With this method we can evaluate the spatial resolution for different track topologies.
The results for the horizontal tracks as a function of the drift distance for different electronics peaking time are presented in Fig.~\ref{fig:SR_vs_Z}.

The dependence of resolution on the drift distance $Z$ is expected to follow:

\begin{equation}
    \sigma(Z)=\sqrt{\sigma_0^2+C^2_{diff}/N_{eff}\times Z}
    \label{eq:SR_vs_Z}
\end{equation}

where $\sigma_0$ is the resolution at null drift distance, $\rm C_{diff}$ is the transversal diffusion constant and $\rm N_{eff}$ is the number of effective electrons~\cite{Colas:2010zz}. The observed dependence is in agreement with this prediction. The peaking time mildly affects the $\rm C^2_{diff}/N_{eff}$ term, but changes the $\sigma_0$. A larger peaking time results in a higher amplitude in the neighbour pads and higher pad multiplicity. Thus we have more robust information for the PRF fit and the track position reconstruction is more precise.

\begin{figure}[H]
    \centering
       \includegraphics[width=0.6\linewidth]{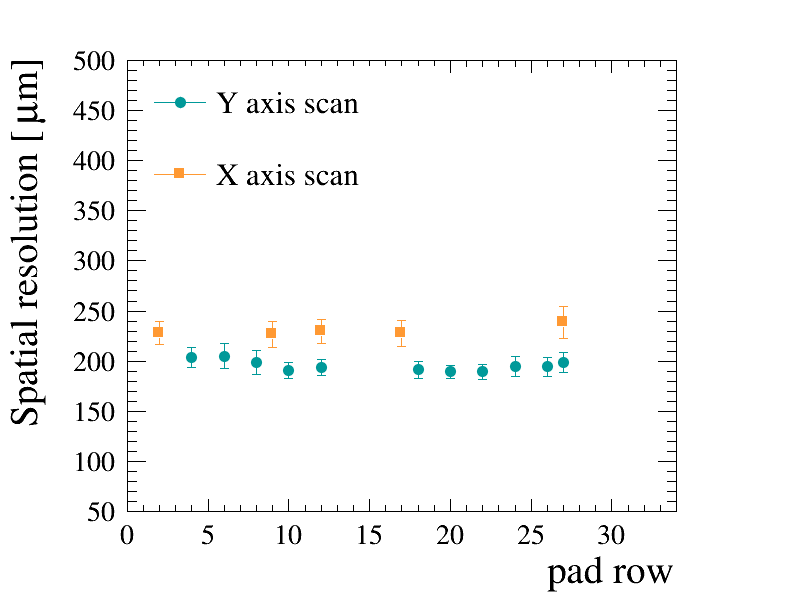}
   \caption{
        Spatial resolution for different Y and X positions at 412 ns peaking time.
    }
    \label{fig:SR_vs_XY}
\end{figure}

The spatial resolution can also depend on the ERAM module characteristics, such as its gain and the local RC value. To investigate these possible dependencies we used a scan done at fixed drift distance but with horizontal tracks crossing the ERAM at different Y positions and (after rotation by 90$^\circ$) vertical tracks crossing the ERAM at different X positions. The spatial resolution obtained for these different X and Y positions is shown in Fig.~\ref{fig:SR_vs_XY}. No large differences are observed indicating that possible local non-uniformities on the ERAM module do not affect the spatial resolution.
The better resolution observed in the Y scan is due to the rectangular shape of the pads that are smaller when different ERAM rows are grouped into clusters.

\subsection{Spatial resolution for inclined tracks}

\begin{figure}[H]
    \centering
    \begin{minipage}{0.48\linewidth}
        \centering
        \includegraphics[width=\linewidth]{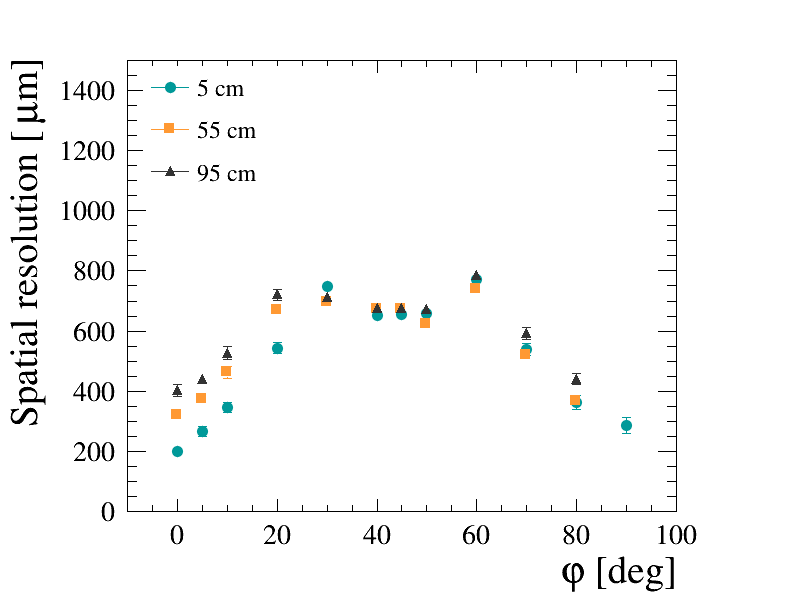} \\ a
    \end{minipage}
    \begin{minipage}{0.48\linewidth}
        \centering
        \includegraphics[width=\linewidth]{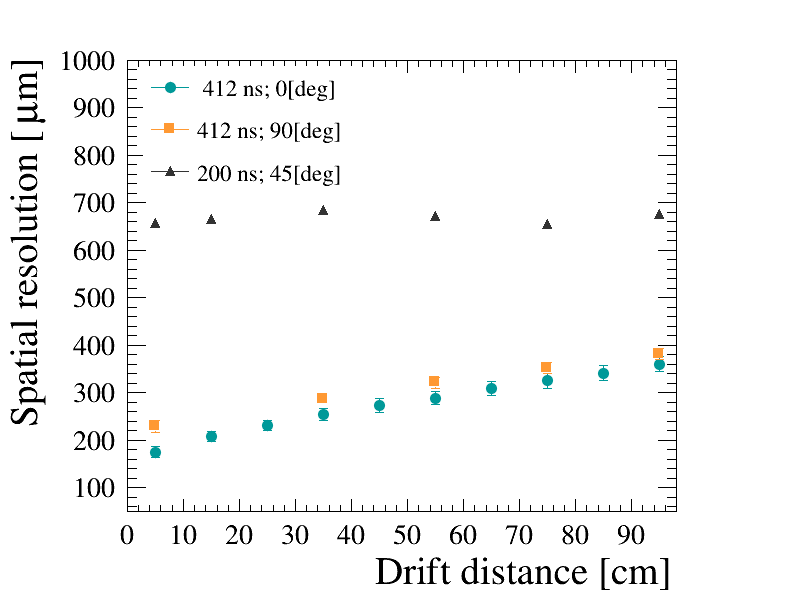} \\ b
    \end{minipage}
    \caption{
        Spatial resolution for different angles of the electron tracks inclination in the pad plane with 200 ns peaking time, 0.2~T magnetic field and for various drift distances: 5~cm, 55~cm, 95~cm (a) and spatial resolution versus drift distance for horizontal (0 deg), inclined (45 deg) and vertical (90 deg) tracks.
    }
    \label{fig:SR_vs_phi}
\end{figure}

The collected test beam data allow studying spatial resolution as a function of the angle of inclination of the tracks with respect to the ERAM module plane. The novelty
with respect to the studies performed in~\cite{Attie:2021yeh} is that we could evaluate the spatial resolution performances for inclined tracks at long drift distances. The results are presented in Fig.~\ref{fig:SR_vs_phi} (a), where for angles from 0 to 30 degrees and from 70 to 90 degrees clusters are defined along the rows/columns (horizontal/vertical fit), and for tracks with the angle between 40 and 60 degrees, the clusters are formed along the diagonals as described in Section \ref{sec:reco}. In~\cite{Attie:2021yeh} it has been shown that the use of a diagonal clustering for highly inclined tracks significantly improves the spatial resolution with respect to the use of horizontal or vertical clustering.

Fig.~\ref{fig:SR_vs_phi}~(a) demonstrates that, while the spatial resolution depends on the angle, it stays between 200 and 800 ${\rm\text{\textmu} m}$ for all the analyzed samples. In particular, it is interesting to notice in Fig.~\ref{fig:SR_vs_phi}~(b) that, while the spatial resolution degrades with the drift distance for horizontal and vertical tracks, it is constant for inclined tracks. 

The behavior for diagonal tracks can be understood considering that the spatial resolution depends on the charge spread over a certain amount of pads (multiplicity). A diagonal clustering algorithm leads to a smaller mean multiplicity than in the case of horizontal/vertical clustering as it was shown in Fig.~\ref{fig:mean_multi}. With the diagonal clustering pad size becomes effectively $\sqrt{2}$ times larger, thus degrading the spatial resolution but also making the effect of the transverse diffusion less significant.

\begin{figure}[!ht]
    \centering
    \begin{minipage}{0.48\linewidth}
        \centering
        \includegraphics[width=\linewidth]{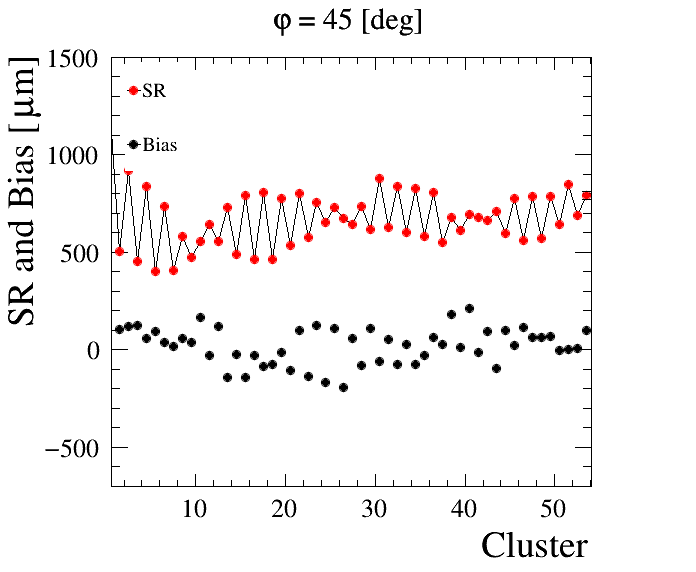} \\ a
    \end{minipage}
    \begin{minipage}{0.48\linewidth}
        \centering
        \includegraphics[width=\linewidth]{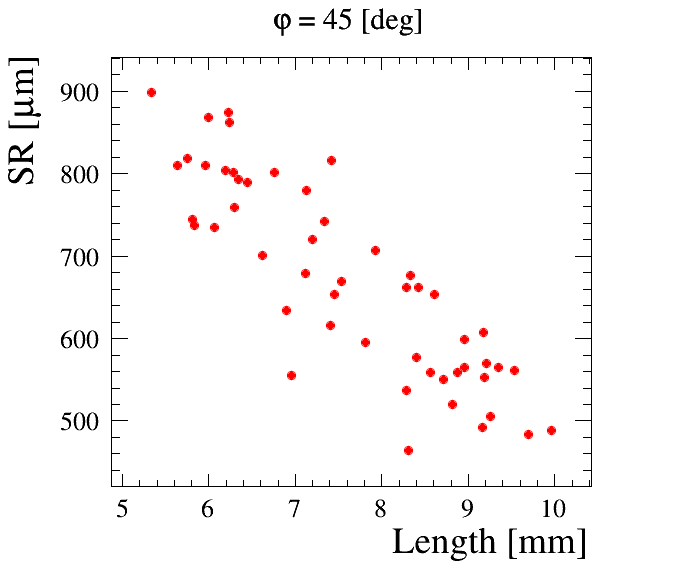} \\ b
    \end{minipage}
    \caption{Tracks of 45 degrees inclination: spatial resolution (SR) per cluster (a) and spatial resolution with respect to mean track segment length in a cluster (b).}
    \label{fig:SR_vs_Length}
\end{figure}

Moreover, diagonal clustering implies a dependence of the resolution on the length of the track in the cluster. This causes an oscillatory behavior in the spatial resolution versus the cluster that is shown in~Fig.~\ref{fig:SR_vs_Length}~(a). Fig.~\ref{fig:SR_vs_Length}~(b) shows the dependence of the spatial resolution on the track length per cluster for inclined tracks. It is clearly seen that, as expected, the resolution improves for longer track segment lengths within the cluster.  

\subsection{Biases in spatial resolution}
\label{sec:sr_biases}

The bias of the track position in each cluster of the ERAM plane is defined as the mean of the Gaussian fit of the distribution of residuals per cluster, and can be referred to as the systematic uncertainty of the track position estimation. In \cite{Attie:2021yeh,https://doi.org/10.48550/arxiv.physics/0510085} it is shown that the biases depend on the track position.
The data collected during this campaign allow for a deeper study of biases, in particular their dependence on drift distance for both horizontal and inclined tracks, as well as their behavior for various magnetic field strengths (see Sec.~\ref{sec:ecrossb}). \par

\begin{figure}[!ht]
    \centering
    \begin{minipage}{0.48\linewidth}
        \centering
        \includegraphics[width=\linewidth]{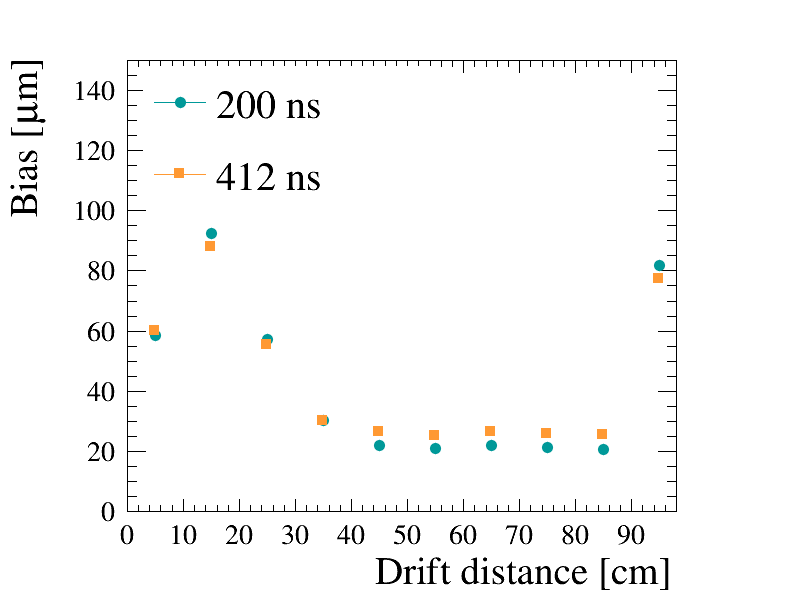} \\ a
    \end{minipage}
    \begin{minipage}{0.48\linewidth}
        \centering
        \includegraphics[width=\linewidth]{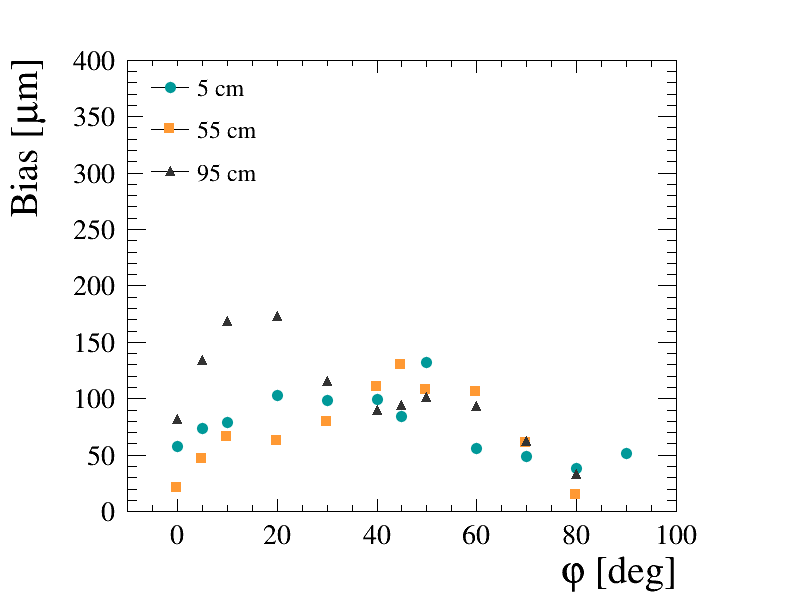} \\ b
    \end{minipage}
    \caption{Track position bias with respect to the drift distance for the horizontal (a) and inclined (b) tracks with a magnetic field of 0.2 T,  and peaking time of 200 ns.}
    \label{fig:SRBiasRatio_vs_Length}
\end{figure}

Fig.~\ref{fig:SRBiasRatio_vs_Length}~(a) shows the bias as a function of the drift distance for horizontal tracks. In this figure, the bias is defined as the mean of the absolute values of the biases per cluster. It is observed that biases are larger at short and long drift distances than at distances corresponding to the middle of the drift volume. \par

For inclined tracks, instead, as shown in  Fig.~\ref{fig:SRBiasRatio_vs_Length}~(b), the biases do not depend on the drift distance or on the $\phi$ angle of the tracks reconstructed using diagonal clustering. \par

\begin{figure}[!ht]
    \centering
    \begin{minipage}{0.48\linewidth}
        \centering
        \includegraphics[width=\linewidth]{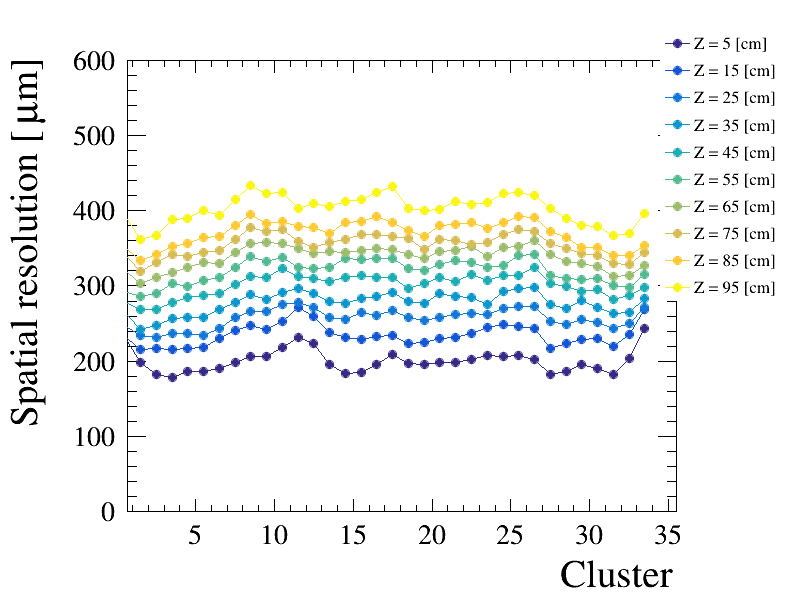} \\ a
    \end{minipage}
    \begin{minipage}{0.48\linewidth}
        \centering
        \includegraphics[width=\linewidth]{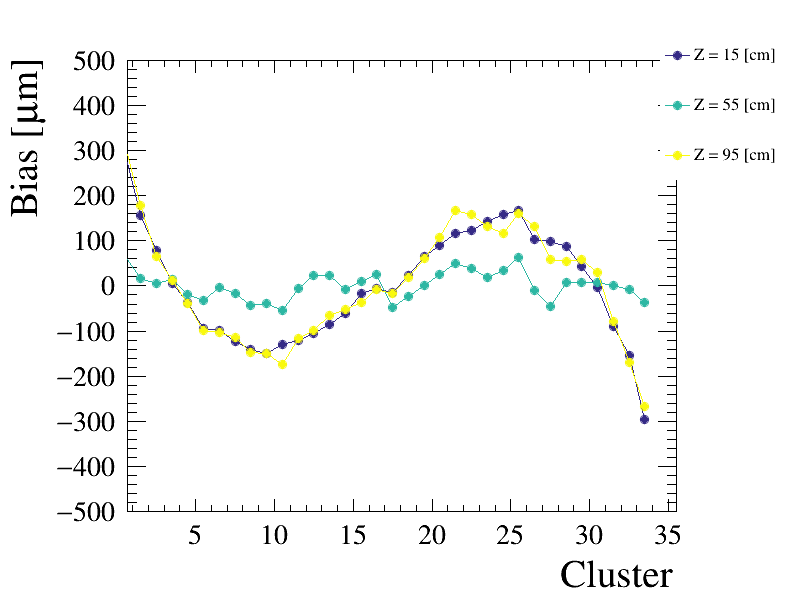} \\ b
    \end{minipage}
    
    \caption{Spatial resolution (a) and track position bias (b) distributions per cluster for horizontal tracks with a magnetic field of 0.2 T, and peaking time of 200 ns at various drift distances.}
    \label{fig:SRratioBias_vs_column}
\end{figure}

In order to further investigate the behavior observed in Fig.~\ref{fig:SRBiasRatio_vs_Length}, the dependence of the spatial resolution and bias per column on the drift distance is shown in Fig.~\ref{fig:SRratioBias_vs_column}. For the smallest and largest drift distances, biases have visible patterns with large and opposite biases at the beginning and at the end of the track. This pattern is not observed for tracks in the center of the ERAM.

The observed dependencies of the biases are not reproduced by the simulation and can point to effects related to non-uniformities in the magnetic field, not accounted for in the simulation, and the E$\times$B effect. In Sect.~\ref{sec:ecrossb} we will discuss the impact of these effects on the observed tracks, together with a study of the observed biases as a function of the applied magnetic field.

For all the configurations the biases are below $200 \; {\rm\text{\textmu} m}$ and their magnitude is small with respect to the spatial resolution for highly inclined tracks.

\section{dE/dx resolution}
\label{sec:dedx}
The other main goal of the HA-TPC is to perform particle identification by measuring the deposited energy per unit length (dE/dx) by charged particles crossing the gas. The TPC particle identification capability will depend on the dE/dx resolution that can be evaluated with the data from this test beam. 
The dE/dx was measured for horizontal and vertical tracks using the track projection on the pad side. 


For inclined tracks reconstructed with diagonal clustering the deposited energy was corrected for the track segment length in each cluster. Such correction accounts for the non-linear dependence of the charge with respect to the track length caused by the charge contribution from the neighbouring clusters for short track lengths. This procedure was described in~\cite{Attie:2021yeh}.\par

In the context of T2K it is particularly important to be able to distinguish electrons and muons. Such an effort is crucial to estimate electron neutrino contamination in the muon neutrino beam and to predict the expected number of un-oscillated electron neutrinos in the far detector. To distinguish electrons and muons the deposited energy resolution better than 10\% is needed in order to achieve a separation between electrons and muons higher than 3 sigma.

The mean deposited energy per unit track length is calculated per track using the truncated mean method already introduced in~\cite{Attie:2021yeh}. 
The method consists in sorting the clusters according to their dE/dx and removing a fraction of those that have the largest energy deposition per unit length.
Such a contribution is caused by fluctuations in the ionization processes and leads to the smearing and to the tail on the right hand side of the dE/dx spectrum. 

The truncation factor is optimized with the data and we found its best value to be 0.7 which is the same as for the DESY 2019 test beam data~\cite{Attie:2021yeh}. This means that 70\% of the clusters are kept for deposited energy per unit length calculations. \par
The dE/dx per cluster was calculated by taking the maximum of the sum of the waveforms of the pads constituting the cluster. Various cluster charge definitions were studied in~\cite{Attie:2021yeh} and it was shown that the charge defined using the sum of the waveforms results in a better dE/dx resolution. For each track sample the resolution was calculated as the ratio of sigma over mean of the Gaussian fit of the corresponding dE/dx distribution.\par

\begin{figure}[H]
    \centering
        \includegraphics[width=0.6\linewidth]{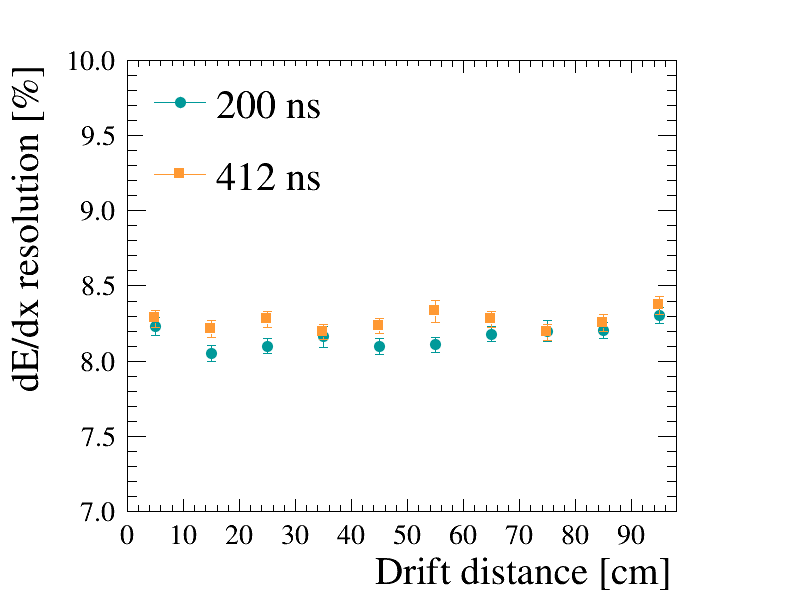} \
    \caption{
        dE/dx resolution with respect to the drift distance for 
        horizontal tracks
        with a magnetic field of 0.2 T, and peaking times of 200 ns and 412 ns.
    }
    \label{fig:dEdx_vs_Z}
\end{figure}

Fig.~\ref{fig:dEdx_vs_Z} shows the dE/dx resolution measured for horizontal tracks for various drift distances.

\begin{figure}[H]
    \centering
        \includegraphics[width=0.6\linewidth]{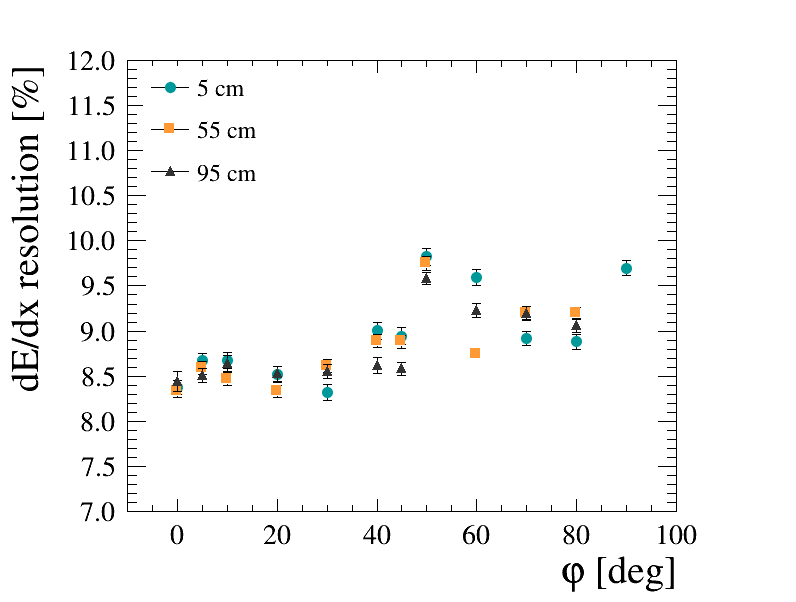} \
    \caption{
        dE/dx resolution versus different inclination angles for 200 ns peaking time, 0.2~T magnetic field and for various drift distances: 5~cm, 55~cm, 95~cm.
    }
    \label{fig:dEdx_vs_phi}
\end{figure}

Fig.~\ref{fig:dEdx_vs_phi} shows the dE/dx resolution at various drift distances as a function of the track inclination angle in the ERAM plane. 

These studies show that the dE/dx resolution is $\sim$8.5\% for horizontal tracks and stays between 7.5\% and 9.6\% for inclined and vertical tracks.  Furthermore, it is independent of drift distance and of the electronics peaking time. It has been observed that the dE/dx resolution is controlled by balancing two factors: the mean charge per cluster and the number of clusters. Fig.~\ref{fig:dEdx_vs_phi}  shows that the dE/dx resolution worsens for the angles $> 45$ degrees since for such angles a smaller number of clusters is reconstructed per track due to the rectangular shape of the ERAM.

\begin{figure}[H]
    \centering
        \includegraphics[width=0.6\linewidth]{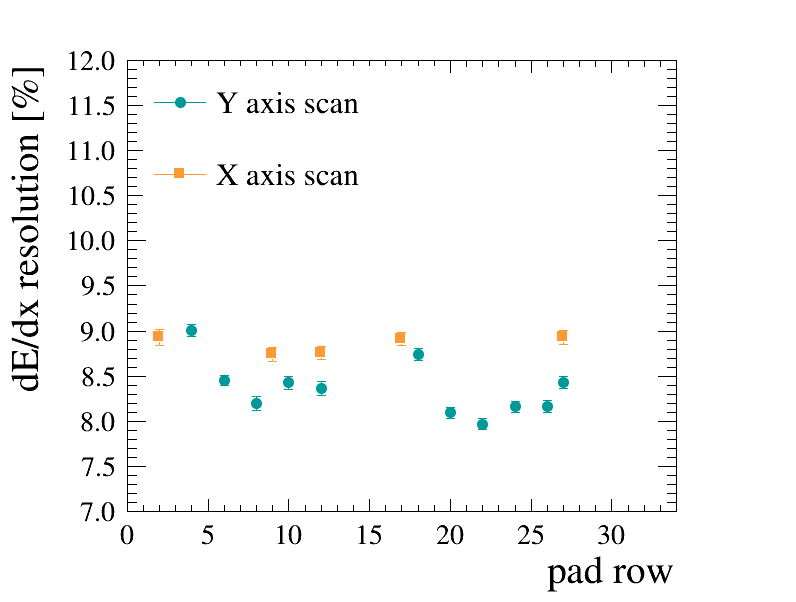}
    \caption{
        dE/dx resolution for tracks entering the ERAM at different X and Y positions.
    }
    \label{fig:dEdx_vs_XY}
\end{figure}

Finally, as in the case of spatial resolution, we looked for effects due to non-uniformities of the ERAM by using the X and Y scans. The results are shown in Fig.~\ref{fig:dEdx_vs_XY}. We observe some differences, possibly due to non-uniformities in the gain of the ERAM, but in general, the resolution is below 10\% for all the scans.

The observed results prove that the ERAM dE/dx resolution fulfills the requirements for the ND280 upgrade.

\section{Comparison between data and simulation}
\label{sec:datasimu}
In this section we compare the performances of the ERAM for spatial and dE/dx resolution between data and simulation.

\begin{figure}[!ht]
    \centering
    \includegraphics[width=0.5\linewidth]{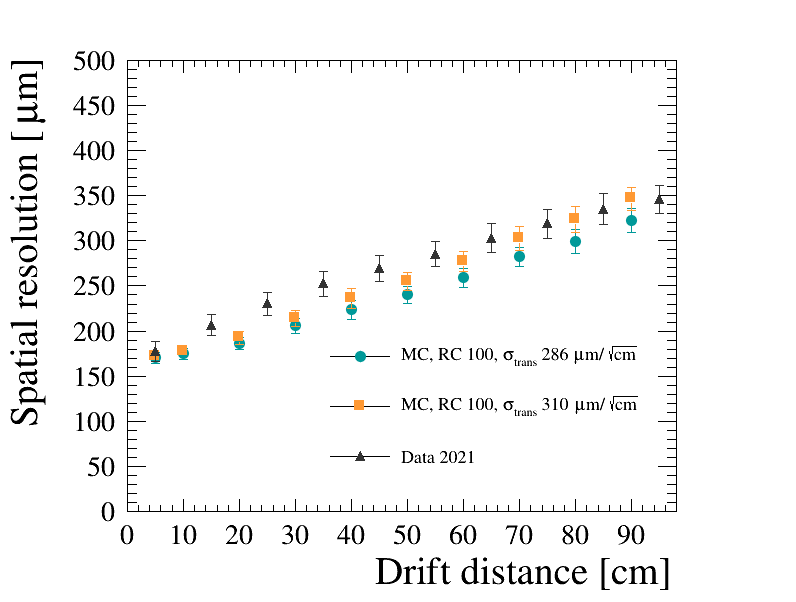}
    
    \caption{Spatial resolution as a function of the drift distance for data at 412 ns peaking time and MC samples generated with different values of transverse diffusion.}
    \label{fig:data_mc_drift}
\end{figure}

Fig.~\ref{fig:data_mc_drift} shows the spatial resolution for the data and MC samples as a function of the drift distance. 
The simulation reproduces better the behaviour observed in the
data when a larger value for the transverse diffusion coefficient $\sigma_{trans}$ is used (when compared to the default value set in the ND280 simulation of the vertical TPCs $\sigma_{trans}=286\ {\rm\text{\textmu} m}/\sqrt{\rm cm}$).


The diffusion can be affected by the magnetic field configuration as well as environmental conditions such as temperature and pressure, or by the amount of oxygen and water contamination in the chamber. 

We made simulations with different values of $\sigma_{trans}$. As expected, in general, increasing $\sigma_{trans}$ results in a worse spatial resolution for large drift distances. 
A satisfactory agreement was found by increasing the transverse diffusion by 8\%, changing it to $\sigma_{trans}=310\ {\rm\text{\textmu} m} /\sqrt{\rm cm}$. Furthermore, while in the data we observe a dependence on the drift distance in good agreement with the one expected from Eq.~\ref{eq:SR_vs_Z}, in the simulation we find that this dependence is linear. The origin of this difference between data and simulation is under investigation. 

\begin{figure}
     \centering
    \begin{minipage}{0.45\linewidth}
        \centering
        \includegraphics[width=\linewidth]{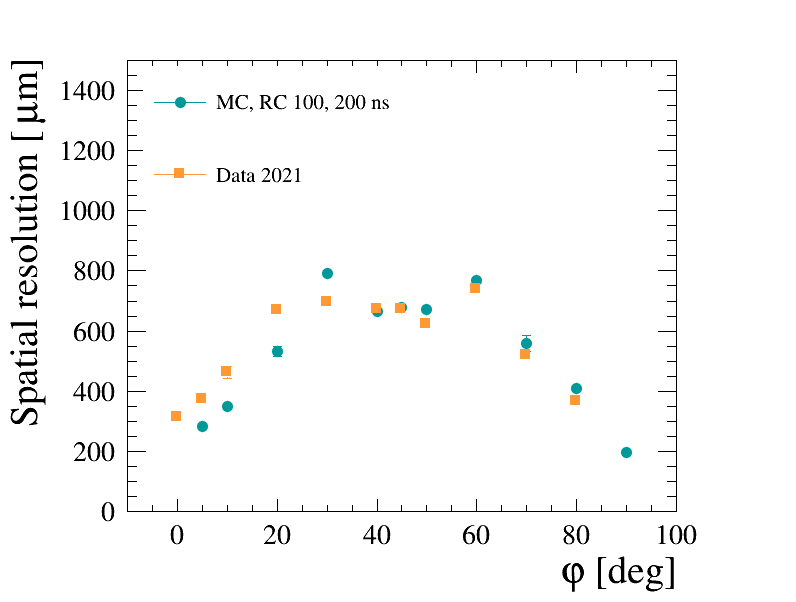} \\ a
    \end{minipage}
    \begin{minipage}{0.49\linewidth}
        \centering
        \includegraphics[width=\linewidth]{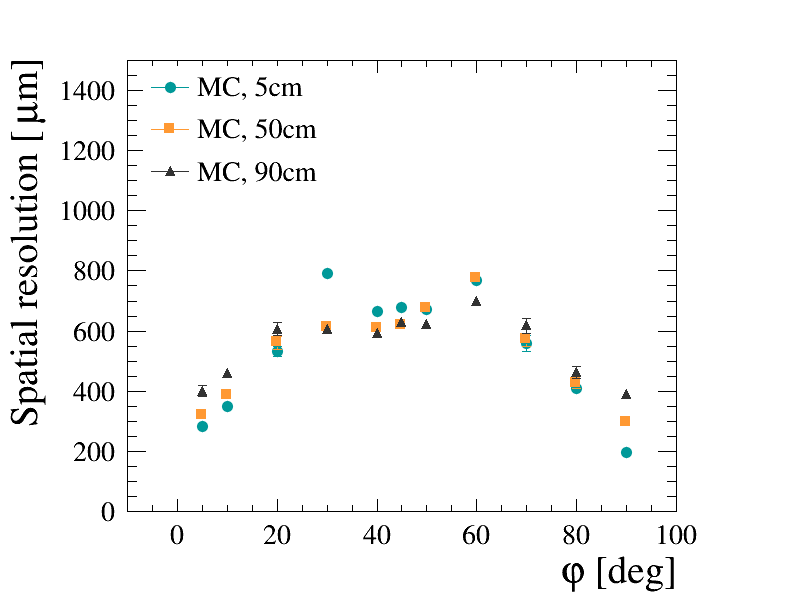} \\ b
    \end{minipage}
    \caption{Spatial resolution as a function of inclination angle for (a) data and MC samples at 200 ns peaking time and 55 cm drift distance and (b) simulation for different drift distances.}
    \label{fig:data_mc_phi}
\end{figure}

In Fig.~\ref{fig:data_mc_phi}~(a) we show the comparison in the spatial resolution between data and simulation for tracks at different angles. 

In Fig.~\ref{fig:SR_vs_phi}~(b) it was found that the spatial resolution weakly depends on the drift distance for the highly inclined tracks. This effect was cross-checked and confirmed with the simulation. Fig.~\ref{fig:data_mc_phi}~(b) shows large effect of the drift distance on the tracks close to $0^\circ$ and $90^\circ$, but much smaller effect at $45^\circ$.

\begin{figure}[!ht]
    \centering
    \begin{minipage}{0.49\linewidth}
        \centering
        \includegraphics[width=\linewidth]{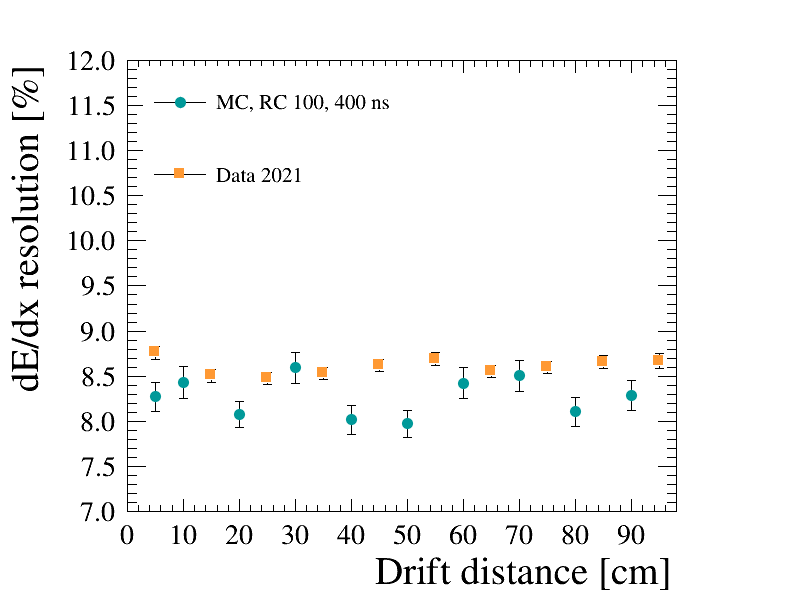} \\ a
    \end{minipage}
    \begin{minipage}{0.49\linewidth}
        \centering
        \includegraphics[width=\linewidth]{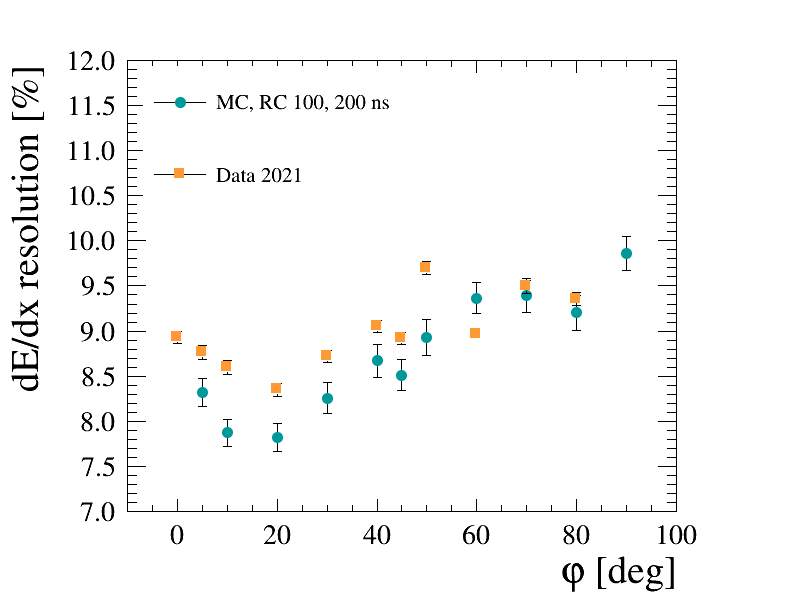} \\ b
    \end{minipage}
    \caption{dE/dx resolution in data and MC for (a) horizontal tracks (at 412 ns peaking time) as a function of the drift distance and for (b) inclined tracks (at 200 ns peaking time) as a function of the track inclination angle. The dE/dx is calculated using the sum of the cluster charge.}
    \label{fig:mc_dedx}
\end{figure}

Concerning the deposited energy resolution, the simulation reproduces reasonably well the data for both horizontal and inclined tracks, as shown in Fig.~\ref{fig:mc_dedx}. The better resolution observed in the simulation could be due to the non-uniformities in the gain that are not introduced in the simulation.

\section{E x B effect}
\label{sec:ecrossb}
Inhomogeneities in the magnetic field can create distortions of the image of the track projected on the ERAM. These distortions are not expected to impact the spatial resolution that is computed based on the track image on the ERAM, but can affect the determination of the reconstructed momentum. This effect is expected to be small in ND280, where the magnetic field inside the magnet has been measured with a dedicated campaign~\cite{Abe:2011ks}, and can be larger in the PCMAG used in DESY, where inclined tracks were observed even for a horizontal beam, as shown in Fig.~\ref{fig:track_profile}.

It is worth saying that this effect cannot be explained by the curvature induced by the magnetic field that is negligible for the operational magnetic field and electron momenta used in the test beam.

Compelling arguments support the hypothesis of ``E $\times$ B effect'' accounting for the inclination of tracks projected on the ERAM.  
Below we present the explanation of the effect. The drift velocity is given by the Langevin equation:

\begin{equation}
    \vec{V_d} = \frac{\mu}{1 + (\omega\tau)^2}\left( \vec{E} + (\omega\tau)\frac{\vec{E}\times\vec{B}}{|\vec{B}|} + (\omega\tau)^2\frac{(\vec{E}\cdot\vec{B})\vec{B}}{|\vec{B}|^2}\right),
\end{equation}

where $\mu = \frac{e}{m}\tau$ is the electron mobility in the gas, $\omega = \frac{eB}{m}$, and $\tau$ is the time between two collisions. 

The drift velocity components are defined as $\vec{V_0} = \frac{\mu}{1 + (\omega\tau)^2} \vec{E}$, $\vec{V_1} =\frac{\mu}{1 + (\omega\tau)^2} \cdot (\omega\tau)\frac{\vec{E}\times\vec{B}}{|\vec{B}|}$ and $\vec{V_2} = \frac{\mu}{1 + (\omega\tau)^2} \cdot (\omega\tau)^2\frac{\vec{B}\cdot(\vec{E}\vec{B})}{|\vec{B}|}$. The angle between the electric and magnetic field is defined as $\delta$ so that $|\vec{E}\times\vec{B}| = EBsin(\delta)$.  

Assuming small $\delta$,  the $\vec{V_2}$ component aligns with $\vec{V_0}$ component. 
Then the drifting electrons will move transversely in the $\vec{V_1}$ direction
and will be projected on the ERAM with a shift $\Delta$:
\begin{equation}
\Delta = Z_{drift}\times \left<\dfrac{v_{y}}{v_{z}}\right> = 
Z_{drift}\times \left( \frac{\left<\delta\right>\omega\tau}{1 + (\omega\tau)^2}\right)
\label{eq:exb_deltay}
\end{equation}
where $Z_{drift}$ is the drift distance and $\left<\delta\right>$ is the average value of $\delta$
along the trajectory of the drifting electron. 

Since $\left<\delta\right>$ varies along the track, an apparent inclination of the track is observed
which can be evaluated with $\phi_{app}$:

\begin{equation}
\phi_{app} = \mathrm{atan}\left(\dfrac{y_L - y_R}{X_{ERAM}}\right) 
\label{eq:exb}
\end{equation}

where $X_{ERAM}$ is the width of the ERAM 
and $y_L$ and  $y_R$ are the track vertical positions on the Left and Right edges.  

\begin{figure}
     \centering
     \includegraphics[width=0.5\textwidth]{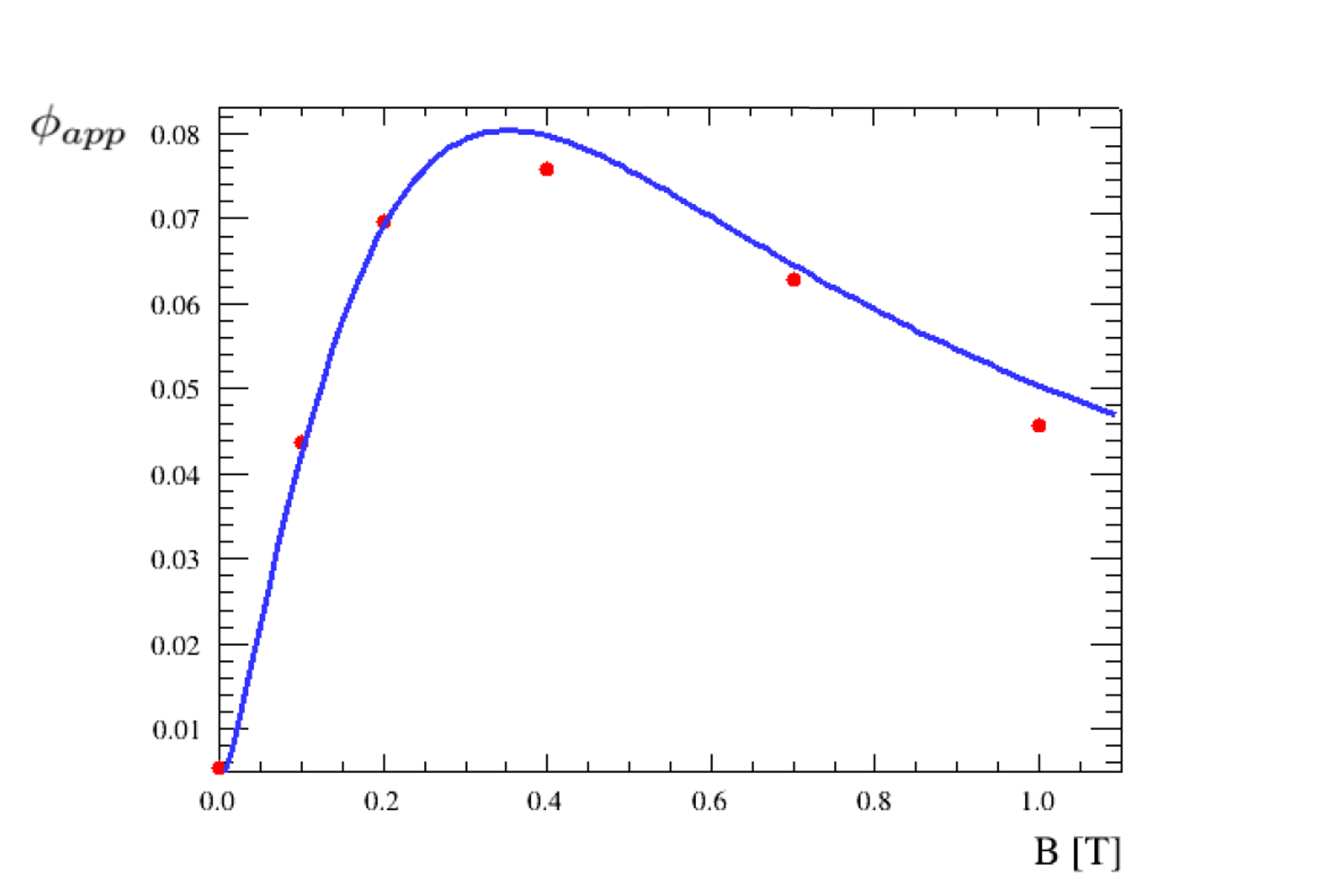}
     \caption{
     The red points show the dependence of observed track inclination on the magnetic field strength. 
     Assuming that the $\delta$ values at each end of the track are opposite,
     these inclinations correspond to twice the displacement given by Eq.~\ref{eq:exb}. 
     The theoretical prediction is shown in blue; the model depends only on $|\delta|$ and the fit finds $|\delta|=0.018$ as the best estimate.
     Notice the slight deviation for high magnetic field, where the bending effect of the longitudinal magnetic field starts to be relevant.
     }
     \label{fig:ExB_differentB}
 \end{figure}

It follows from Eq.~(\ref{eq:exb_deltay}) that the apparent inclination due to E$\times$B effect is maximum for $\omega\tau=\mu B=1$. In our case, where the electron mobility is expected around $\mu=$2.8~$\rm T^{-1}$, this means B$\simeq$0.36~T, which is consistent with what Fig.~\ref{fig:ExB_differentB} shows.  

 As it can be seen in Fig.~\ref{fig:track_profile}, the electrons drifting to the leftmost region of the ERAM are shifted upwards, while the ones drifting to the rightmost region are shifted downwards.
The displacements perpendicular to the track depend on the component of the magnetic field on the track direction. 
 It turns out that this component flips sign when the track crosses the inner volume of 
 the PCMAG solenoid due to the symmetry of the radial components of its field.

 To study this effect, we simulated the motion of drifting electrons with Garfield++~\cite{garfield}, under the proper electric and magnetic field conditions and the gas mixture used in the detector. For the magnetic field, we use the map based on previous measurements at DESY~\cite{PCMAG-map}. 

\begin{figure}
    \centering
    \includegraphics[width=0.5\textwidth,height=4.4cm]{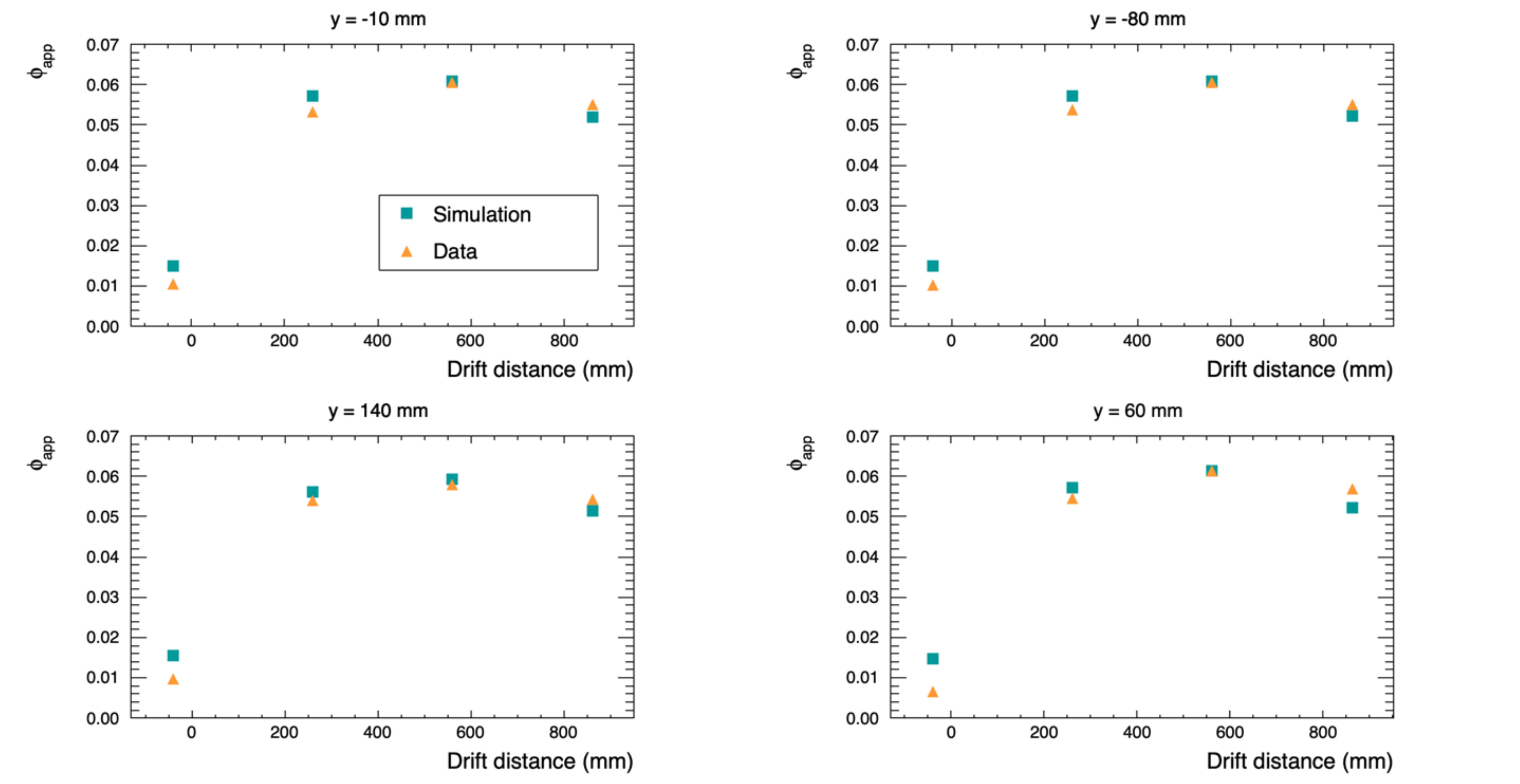}
    \caption{Track inclinations in data and simulations for B~=~0.2~T. The simulation uses Garfield++~\cite{garfield} as described in the text.}
    \label{fig:yz-distortion}
\end{figure}

From both data and simulation, we compute the inclination of tracks projected on the ERAM, and compare the results. Fig.~\ref{fig:yz-distortion} shows that the simulation is able to reproduce the vertical displacement of drifting electrons as observed in the data. 

\begin{figure}[H]
    \centering
    \begin{minipage}{0.48\linewidth}
        \centering
        \includegraphics[width=\linewidth]{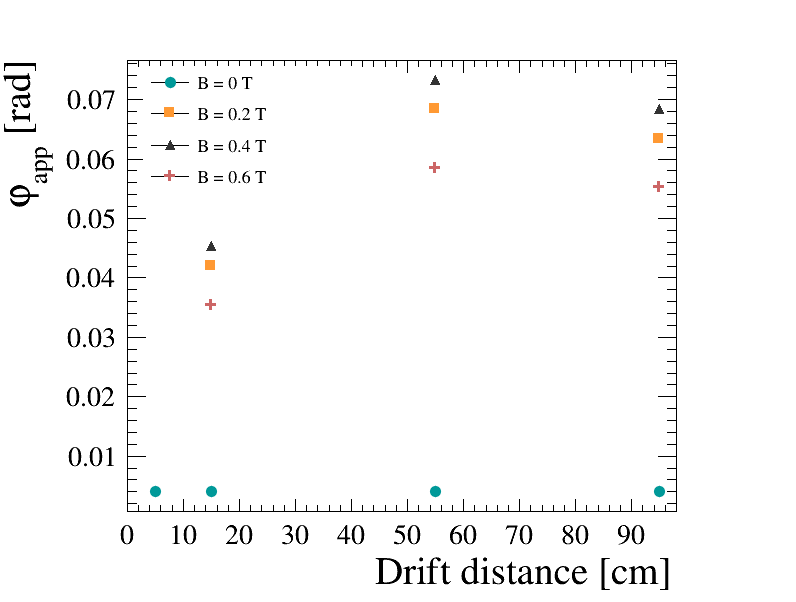} \\ a
    \end{minipage}
    \begin{minipage}{0.48\linewidth}
        \centering
        \includegraphics[width=\linewidth]{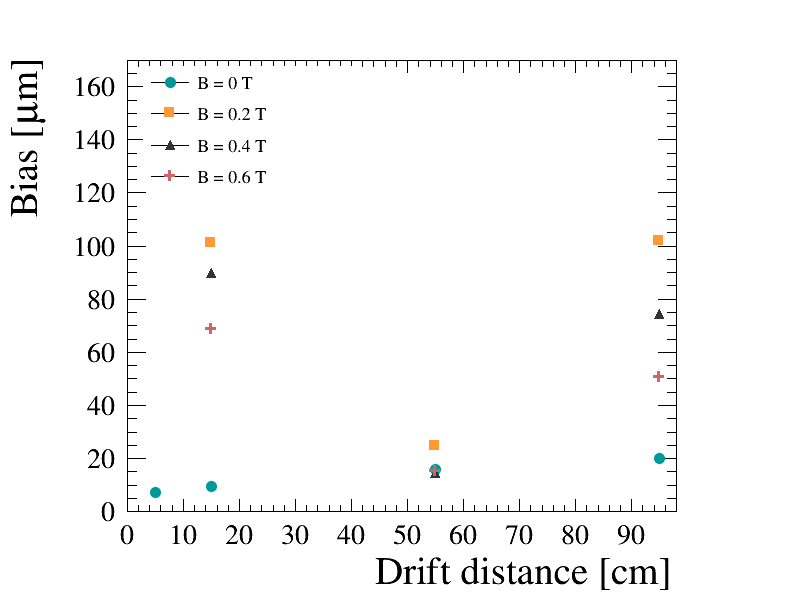} \\ b
    \end{minipage}
    \caption{
       Track inclinations (a) and track position bias (b) with respect to the drift distance for horizontal tracks for different values of magnetic field at the peaking time of 200 ns.
    }
    \label{fig:Deltay_vs_Z}
\end{figure}

The E$\times$B effect dependence on the magnetic field is shown in Fig.~\ref{fig:Deltay_vs_Z}~(a)
for different values of magnetic field and different drift distances. 
As expected, without the magnetic field, there is no displacement in Y. When the magnetic field is on, as expected, a larger displacement is observed for small values of the magnetic field.

In Fig.~\ref{fig:Deltay_vs_Z} (b) we show the biases in the spatial resolution for different values of the magnetic field and as a function of the drift distance. The biases are small for all the drift distances when the magnetic field is off. When the magnetic field is on we observe larger biases for short and long drift distances, similar to the ones shown in Fig.~\ref{fig:SRratioBias_vs_column}. The larger biases are observed for values of the magnetic field of 0.2 and 0.4~T where the E$\times$B effects are the largest. 

The E$\times$B effect generates displacements of the charge landing position on the anode surface
which result in the observed inclination of the tracks.
However, these displacements do not scale linearly along the track and deviations from the inclined direction are expected which could explain the biases observed in the spatial resolution described in Sect.~\ref{sec:spatial}. 
Further investigations are ongoing on this point but it is already clear that 
these biases share the same behavior with respect to the magnetic field as the E$\times$B effect reported above.

\section{Discussion and further  improvements}
\label{sec:discussion}


The ERAM ``charge spreading'' effect has only a small impact on the resolution of the measurement of the dE/dx. In this approach, as also shown in~\cite{Attie:2021yeh}, it is important to include in the dE/dx computation the charges ``shared'' among pads due to the diffusion and separate it from the ``charge spreading'' time-dependent effect.

In contrast, the ERAM ``charge spreading'' effect allows for a significant improvement of the spatial resolution, as the obtained results amply illustrate.
Since both spatial and dE/dx resolutions depend on the relative gain between neighbouring pads, 
gain of each pad is being measured on a dedicated setup and accounting for pad-to-pad variations could potentially lead to better performances.

That is why measurements of RC and gain maps for all produced ERAM modules are being performed using a point-like iron X-ray source placed close to the ERAM. These measurements are complementary to the test beams data analyses and they are described in details in~\cite{Attie:2023uqx}.

The newly developed simulation provides a reasonable description of both the ``charge sharing'' and ``charge spreading'' effects introduced in Sec.~\ref{sec:ERAMtechnology}. This tool allows to understand deeper the data from the test beam, and to disentangle different features, especially when associated with the simpler data from the X-ray surveys.


\section{Conclusions}
\label{sec:conclusion}
In this paper we present the performances of the prototype of the HA-TPCs for the T2K Near Detector upgrade obtained during a test beam at DESY performed in 2021. 
The TPC was instrumented with one of the ERAM detectors that will be installed in the HA-TPCs and the final HA-TPC readout electronics chain.

The test beam data allowed spatial and dE/dx resolutions to be determined as a function of the angle of the track with respect to the ERAM plane for all the drift distances of interest for T2K.
Spatial resolution better than 800 $\text{\textmu}\textrm{m}$ is obtained for all the angles and all the drift distances using a dedicated clustering algorithm which is adapted to the track angle.  The dE/dx resolution better than 10\% is obtained for all the angles.  

The data are compared with a simulation of the ERAM response, including the features of the resistive layer. As shown in this paper, the simulation is able to satisfactory reproduce the observed charge sharing between neighboring pads. Spatial resolution and dE/dx resolution are also in good agreement between data and simulation.

\section*{Acknowledgements}
\begin{sloppypar}
The measurements leading to these results have been performed at the Test Beam Facility at DESY Hamburg (Germany), a member of the Helmholtz Association.
The authors would like to thank the technical team at the DESY II accelerator and test beam facility for the smooth operation of the test beam and the support during the test beam campaign. We also thank the T2K collaboration for the extensive use of the ND280 software.\\
We acknowledge the support of CEA and CNRS/IN2P3, France; DFG, Germany; INFN, Italy; National Science Centre (NCN) and Ministry of Science and Higher Education (Grant No. DIR/WK/2017/05), Poland;   the Spanish Ministerio de Econom\'{i}a y Competitividad  (SEIDI - MINECO) under Grant No.~PID2019-107564GB-I00 (IFAE, Spain). 
IFAE is partially funded by the CERCA program of the Generalitat de Catalunya.\\
In addition, the participation of individual researchers and institutions has been further supported by H2020 Grant No. RISE-GA822070-JENNIFER2 2020, MSCA-COFUND-2016 No.754496, ANR-19-CE31-0001, RFBR grants \#19-32-90100, the Secretariat for Universities and Research of the Ministry of Business and Knowledge of the Government of Catalonia and the European Social Fund (2022FI\_B 00336) and from the program Plan de Doctorados Industriales of the Research and Universities Department of the Catalan Government (2022 DI 011).
\end{sloppypar}





\bibliographystyle{elsarticle-num}
\bibliography{bibliography}

\begin{thebibliography}{10}
\expandafter\ifx\csname url\endcsname\relax
  \def\url#1{\texttt{#1}}\fi
\expandafter\ifx\csname urlprefix\endcsname\relax\def\urlprefix{URL }\fi
\expandafter\ifx\csname href\endcsname\relax
  \def\href#1#2{#2} \def\path#1{#1}\fi

\bibitem{Abe:2011ks}
K.~Abe, et~al., {The T2K Experiment}, Nucl. Instrum. Meth. A659 (2011)
  106--135.
\newblock \href {http://arxiv.org/abs/1106.1238} {\path{arXiv:1106.1238}},
  \href {https://doi.org/10.1016/j.nima.2011.06.067}
  {\path{doi:10.1016/j.nima.2011.06.067}}.

\bibitem{Abe:2011sj}
K.~Abe, et~al., {Indication of Electron Neutrino Appearance from an
  Accelerator-produced Off-axis Muon Neutrino Beam}, Phys. Rev. Lett. 107
  (2011) 041801.
\newblock \href {http://arxiv.org/abs/1106.2822} {\path{arXiv:1106.2822}},
  \href {https://doi.org/10.1103/PhysRevLett.107.041801}
  {\path{doi:10.1103/PhysRevLett.107.041801}}.

\bibitem{Abe:2013hdq}
K.~Abe, et~al., {Observation of Electron Neutrino Appearance in a Muon Neutrino
  Beam}, Phys. Rev. Lett. 112 (2014) 061802.
\newblock \href {http://arxiv.org/abs/1311.4750} {\path{arXiv:1311.4750}},
  \href {https://doi.org/10.1103/PhysRevLett.112.061802}
  {\path{doi:10.1103/PhysRevLett.112.061802}}.

\bibitem{Abe:2019vii}
K.~Abe, et~al., {Constraint on the matter\textendash{}antimatter
  symmetry-violating phase in neutrino oscillations}, Nature 580~(7803) (2020)
  339--344, [Erratum: Nature 583, E16 (2020)].
\newblock \href {http://arxiv.org/abs/1910.03887} {\path{arXiv:1910.03887}},
  \href {https://doi.org/10.1038/s41586-020-2177-0}
  {\path{doi:10.1038/s41586-020-2177-0}}.

\bibitem{T2K:2019eao}
K.~Abe, et~al., {J-PARC Neutrino Beamline Upgrade Technical Design Report} (8
  2019).
\newblock \href {http://arxiv.org/abs/1908.05141} {\path{arXiv:1908.05141}}.

\bibitem{T2K:2019bbb}
K.~Abe, et~al., {T2K ND280 Upgrade - Technical Design Report} (1 2019).
\newblock \href {http://arxiv.org/abs/1901.03750} {\path{arXiv:1901.03750}}.

\bibitem{T2KND280FGD:2012umz}
P.~A. Amaudruz, et~al., {The T2K Fine-Grained Detectors}, Nucl. Instrum. Meth.
  A 696 (2012) 1--31.
\newblock \href {http://arxiv.org/abs/1204.3666} {\path{arXiv:1204.3666}},
  \href {https://doi.org/10.1016/j.nima.2012.08.020}
  {\path{doi:10.1016/j.nima.2012.08.020}}.

\bibitem{T2KND280TPC:2010nnd}
N.~Abgrall, et~al., {Time Projection Chambers for the T2K Near Detectors},
  Nucl. Instrum. Meth. A 637 (2011) 25--46.
\newblock \href {http://arxiv.org/abs/1012.0865} {\path{arXiv:1012.0865}},
  \href {https://doi.org/10.1016/j.nima.2011.02.036}
  {\path{doi:10.1016/j.nima.2011.02.036}}.

\bibitem{Giomataris:2004aa}
I.~Giomataris, R.~De~Oliveira, S.~Andriamonje, S.~Aune, G.~Charpak, P.~Colas,
  A.~Giganon, P.~Rebourgeard, P.~Salin, {Micromegas in a bulk}, Nucl. Instrum.
  Meth. A560 (2006) 405--408.
\newblock \href {http://arxiv.org/abs/physics/0501003}
  {\path{arXiv:physics/0501003}}, \href
  {https://doi.org/10.1016/j.nima.2005.12.222}
  {\path{doi:10.1016/j.nima.2005.12.222}}.

\bibitem{T2K:2021xwb}
K.~Abe, et~al., {Improved constraints on neutrino mixing from the T2K
  experiment with $\mathbf{3.13\times10^{21}}$ protons on target}, Phys. Rev. D
  103~(11) (2021) 112008.
\newblock \href {http://arxiv.org/abs/2101.03779} {\path{arXiv:2101.03779}},
  \href {https://doi.org/10.1103/PhysRevD.103.112008}
  {\path{doi:10.1103/PhysRevD.103.112008}}.

\bibitem{Abe:2019whr}
K.~Abe, et~al., {T2K ND280 Upgrade - Technical Design Report} (2019).
\newblock \href {http://arxiv.org/abs/1901.03750} {\path{arXiv:1901.03750}}.

\bibitem{Dolan:2021hbw}
S.~Dolan, et~al., {Sensitivity of the upgraded T2K Near Detector to constrain
  neutrino and antineutrino interactions with no mesons in the final state by
  exploiting nucleon-lepton correlations}, Phys. Rev. D 105~(3) (2022) 032010.
\newblock \href {http://arxiv.org/abs/2108.11779} {\path{arXiv:2108.11779}},
  \href {https://doi.org/10.1103/PhysRevD.105.032010}
  {\path{doi:10.1103/PhysRevD.105.032010}}.

\bibitem{Assylbekov:2011sh}
S.~Assylbekov, et~al., {The T2K ND280 Off-Axis Pi-Zero Detector}, Nucl.
  Instrum. Meth. A686 (2012) 48--63.
\newblock \href {http://arxiv.org/abs/1111.5030} {\path{arXiv:1111.5030}},
  \href {https://doi.org/10.1016/j.nima.2012.05.028}
  {\path{doi:10.1016/j.nima.2012.05.028}}.

\bibitem{Blondel:2020hml}
A.~Blondel, et~al., {The SuperFGD Prototype Charged Particle Beam Tests}, JINST
  15~(12) (2020) P12003.
\newblock \href {http://arxiv.org/abs/2008.08861} {\path{arXiv:2008.08861}},
  \href {https://doi.org/10.1088/1748-0221/15/12/P12003}
  {\path{doi:10.1088/1748-0221/15/12/P12003}}.

\bibitem{Korzenev:2021mny}
A.~Korzenev, et~al., {A 4\ensuremath{\pi} time-of-flight detector for the
  ND280/T2K upgrade}, JINST 17~(01) (2022) P01016.
\newblock \href {http://arxiv.org/abs/2109.03078} {\path{arXiv:2109.03078}},
  \href {https://doi.org/10.1088/1748-0221/17/01/P01016}
  {\path{doi:10.1088/1748-0221/17/01/P01016}}.

\bibitem{Attie:2687703}
D.~Attié, et~al., \href{http://cds.cern.ch/record/2687703}{{Performances of a
  resistive MicroMegas module for the Time Projection Chambers of the T2K Near
  Detector upgrade}}, Nucl. Instrum. Methods Phys. Res., A
  957~(arXiv:1907.07060) (2019) 163286. 13 p, 26 figures.
\newblock \href {https://doi.org/10.1016/j.nima.2019.163286}
  {\path{doi:10.1016/j.nima.2019.163286}}.
\newline\urlprefix\url{http://cds.cern.ch/record/2687703}

\bibitem{Baron:2008zza}
P.~Baron, D.~Calvet, E.~Delagnes, X.~de~la Broise, A.~Delbart, F.~Druillole,
  E.~Monmarthe, E.~Mazzucato, F.~Pierre, M.~Zito, {AFTER, an ASIC for the
  readout of the large T2K time projection chambers}, IEEE Trans. Nucl. Sci. 55
  (2008) 1744--1752.
\newblock \href {https://doi.org/10.1109/TNS.2008.924067}
  {\path{doi:10.1109/TNS.2008.924067}}.

\bibitem{Calvet:2018lac}
D.~Calvet, {Back-End Electronics Based on an Asymmetric Network for Low
  Background and Medium- Scale Physics Experiments}, IEEE Trans. Nucl. Sci.
  66~(7) (2018) 998--1006.
\newblock \href {http://arxiv.org/abs/1806.07618} {\path{arXiv:1806.07618}},
  \href {https://doi.org/10.1109/TNS.2018.2884617}
  {\path{doi:10.1109/TNS.2018.2884617}}.

\bibitem{MIDAS}
\href{https://midas.triumf.ca/}{{MIDAS} -- modern data acquisition system}
  (2020).
\newline\urlprefix\url{https://midas.triumf.ca/}

\bibitem{Diener:2018qap}
R.~Diener, et~al., {The DESY II Test Beam Facility}, Nucl. Instrum. Meth. A 922
  (2019) 265--286.
\newblock \href {http://arxiv.org/abs/1807.09328} {\path{arXiv:1807.09328}},
  \href {https://doi.org/10.1016/j.nima.2018.11.133}
  {\path{doi:10.1016/j.nima.2018.11.133}}.

\bibitem{Attie:2019hua}
D.~Atti\'e, et~al., {Performances of a resistive Micromegas module for the Time
  Projection Chambers of the T2K Near Detector upgrade}, Nucl. Instrum. Meth. A
  957 (2020) 163286.
\newblock \href {http://arxiv.org/abs/1907.07060} {\path{arXiv:1907.07060}},
  \href {https://doi.org/10.1016/j.nima.2019.163286}
  {\path{doi:10.1016/j.nima.2019.163286}}.

\bibitem{Attie:2021yeh}
D.~Atti\'e, et~al., {Characterization of resistive Micromegas detectors for the
  upgrade of the T2K Near Detector Time Projection Chambers}, Nucl. Instrum.
  Meth. A 1025 (2022) 166109.
\newblock \href {http://arxiv.org/abs/2106.12634} {\path{arXiv:2106.12634}},
  \href {https://doi.org/10.1016/j.nima.2021.166109}
  {\path{doi:10.1016/j.nima.2021.166109}}.

\bibitem{Attie:2011zz}
D.~Attie, {Beam tests of Micromegas LC-TPC large prototype}, JINST 6 (2011)
  C01007.
\newblock \href {https://doi.org/10.1088/1748-0221/6/01/C01007}
  {\path{doi:10.1088/1748-0221/6/01/C01007}}.

\bibitem{Abgrall:2010hi}
N.~Abgrall, et~al., {Time Projection Chambers for the T2K Near Detectors},
  Nucl. Instrum. Meth. A637 (2011) 25--46.
\newblock \href {http://arxiv.org/abs/1012.0865} {\path{arXiv:1012.0865}},
  \href {https://doi.org/10.1016/j.nima.2011.02.036}
  {\path{doi:10.1016/j.nima.2011.02.036}}.

\bibitem{OUESTRONIC}
{OUESTRONIC} company, \url{https://www.ouestronic.fr/} (2022).

\bibitem{GEANT4:2002zbu}
S.~Agostinelli, et~al., {GEANT4--a simulation toolkit}, Nucl. Instrum. Meth. A
  506 (2003) 250--303.
\newblock \href {https://doi.org/10.1016/S0168-9002(03)01368-8}
  {\path{doi:10.1016/S0168-9002(03)01368-8}}.

\bibitem{Apostolakis2000}
J.~Apostolakis, S.~Giani, L.~Urban, M.~Maire, A.~V. Bagulya, V.~M. Grichine, An
  implementation of ionisation energy loss in very thin absorbers for the
  geant4 simulation package, Nuclear Instruments and Methods in Physics
  Research Section A: Accelerators, Spectrometers, Detectors and Associated
  Equipment 453 (2000) 597--605.
\newblock \href {https://doi.org/10.1016/S0168-9002(00)00457-5}
  {\path{doi:10.1016/S0168-9002(00)00457-5}}.

\bibitem{Dixit:2006ge}
M.~S. Dixit, A.~Rankin, {Simulating the charge dispersion phenomena in micro
  pattern gas detectors with a resistive anode}, Nucl. Instrum. Meth. A566
  (2006) 281--285.
\newblock \href {http://arxiv.org/abs/physics/0605121}
  {\path{arXiv:physics/0605121}}, \href
  {https://doi.org/10.1016/j.nima.2006.06.050}
  {\path{doi:10.1016/j.nima.2006.06.050}}.

\bibitem{Attie:2023uqx}
D.~Attie, et~al., {Characterization of Charge Spreading and Gain of
  Encapsulated Resistive Micromegas Detectors for the Upgrade of the T2K Near
  Detector Time Projection Chambers} (3 2023).
\newblock \href {http://arxiv.org/abs/2303.04481} {\path{arXiv:2303.04481}}.

\bibitem{Ester96adensity-based}
M.~Ester, H.-P. Kriegel, J.~Sander, X.~Xu, A density-based algorithm for
  discovering clusters in large spatial databases with noise, in: Proceedings
  of the Second International Conference on Knowledge Discovery and Data
  Mining, AAAI Press, 1996, pp. 226--231.

\bibitem{GLUCKSTERN1963381}
R.~Gluckstern,
  \href{https://www.sciencedirect.com/science/article/pii/0029554X63903471}{Uncertainties
  in track momentum and direction, due to multiple scattering and measurement
  errors}, Nuclear Instruments and Methods 24 (1963) 381--389.
\newblock \href {https://doi.org/https://doi.org/10.1016/0029-554X(63)90347-1}
  {\path{doi:https://doi.org/10.1016/0029-554X(63)90347-1}}.
\newline\urlprefix\url{https://www.sciencedirect.com/science/article/pii/0029554X63903471}

\bibitem{Boudjemline:2006hf}
K.~Boudjemline, M.~S. Dixit, J.~P. Martin, K.~Sachs, {Spatial resolution of a
  GEM readout TPC using the charge dispersion signal}, Nucl. Instrum. Meth.
  A574 (2007) 22--27.
\newblock \href {http://arxiv.org/abs/physics/0610232}
  {\path{arXiv:physics/0610232}}, \href
  {https://doi.org/10.1016/j.nima.2007.01.017}
  {\path{doi:10.1016/j.nima.2007.01.017}}.

\bibitem{Colas:2010zz}
P.~Colas, {First test results from a Micromegas large TPC prototype}, Nucl.
  Instrum. Meth. A623 (2010) 100--101.
\newblock \href {https://doi.org/10.1016/j.nima.2010.02.161}
  {\path{doi:10.1016/j.nima.2010.02.161}}.

\bibitem{https://doi.org/10.48550/arxiv.physics/0510085}
A.~Bellerive, K.~Boudjemline, R.~Carnegie, M.~Dixit, J.~Miyamoto, E.~Neuheimer,
  A.~Rankin, E.~Rollin, K.~Sachs, J.~P. Martin, V.~Lepeltier, P.~Colas,
  A.~Giganon, I.~Giomataris,
  \href{https://arxiv.org/abs/physics/0510085}{Spatial resolution of a
  micromegas-tpc using the charge dispersion signal} (2005).
\newblock \href {https://doi.org/10.48550/ARXIV.PHYSICS/0510085}
  {\path{doi:10.48550/ARXIV.PHYSICS/0510085}}.
\newline\urlprefix\url{https://arxiv.org/abs/physics/0510085}

\bibitem{garfield}
https://garfield.web.cern.ch/garfield (2010).

\bibitem{PCMAG-map}
C.~Grefe, {Magnetic field map for a large TPC prototype}, Other thesis (6
  2008).
\newblock \href {https://doi.org/10.3204/DESY-THESIS-2008-052}
  {\path{doi:10.3204/DESY-THESIS-2008-052}}.

\end{thebibliography}

\end{document}